\documentclass[useAMS]{mn2e}
\usepackage{graphics}
\usepackage{graphicx}
\usepackage{epsf}
\usepackage{url}
\newif\ifAMStwofonts
\AMStwofontstrue

\newcommand{\dmsqr}{
    \begin{figure}
    \begin{center}
    \leavevmode
    \epsfxsize=\columnwidth    
    \epsfbox{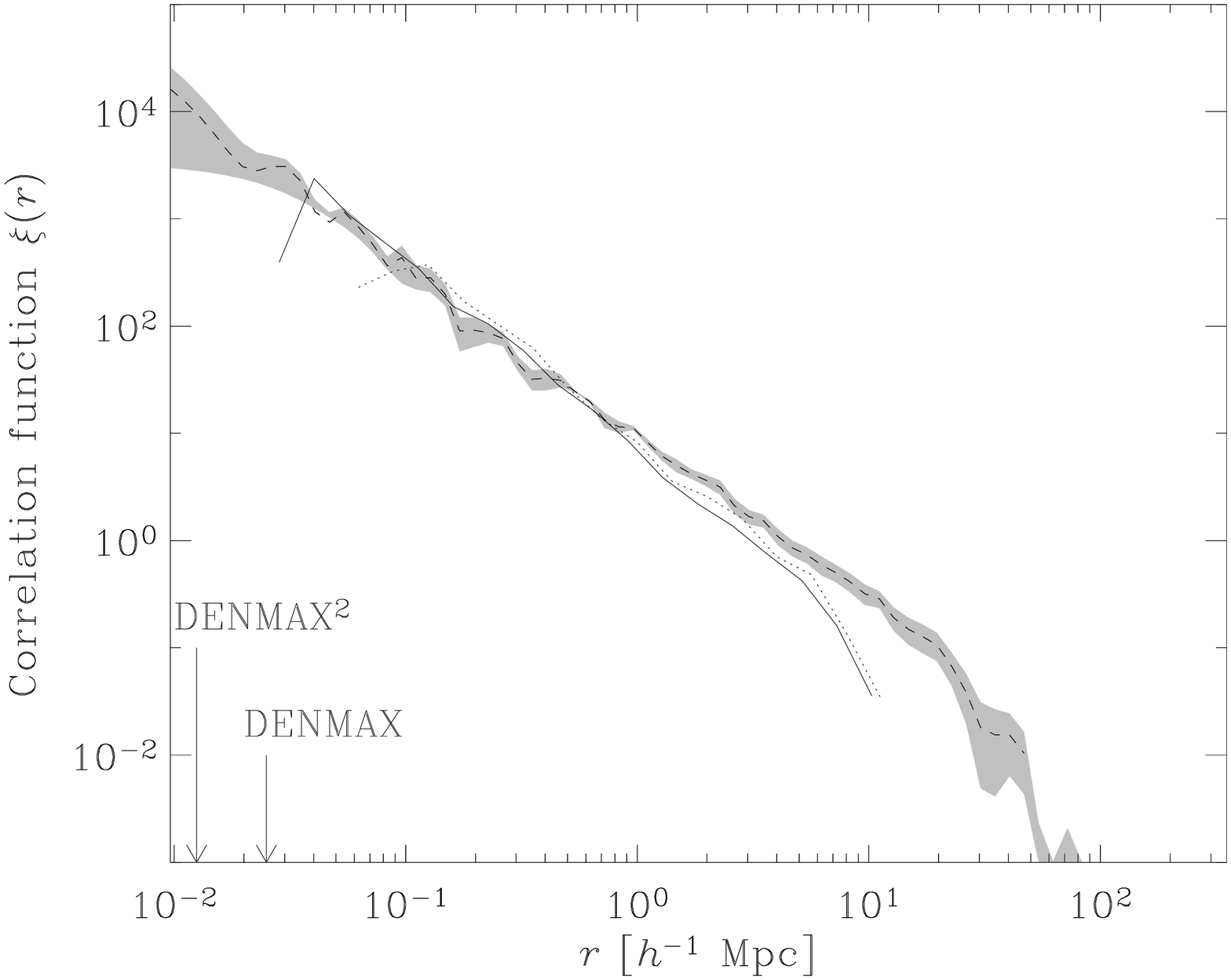}
    \end{center}
    \caption[1]{ \small The effect of added DENMAX$^2$ haloes on the
    correlation function in the 32 $h^{-1}\!$ Mpc simulation.  The dashed
    curve is the PSC$z$ correlation function, and the grey band shows its
    errors.  The dotted curve is the best fit using just DENMAX haloes, and
    the solid curve is the best fit including DENMAX$^2$ haloes.  The
    best-fitting central density cutoff was different in each case: for
    DENMAX, $\rho_{c,min}$ was 225 particles; for DENMAX$^2$, it was 122
    particles.  The arrows show the DENMAX and DENMAX$^2$ smoothing
    lengths: 0.025 and 0.0125 $h^{-1}$ Mpc, respectively.
    \label{dmsqr}
    }
    \end{figure}
}
\newcommand{\mvrho}{
    \begin{figure}
    \begin{center}
    \leavevmode
    \epsfxsize=\columnwidth    
    \epsfbox{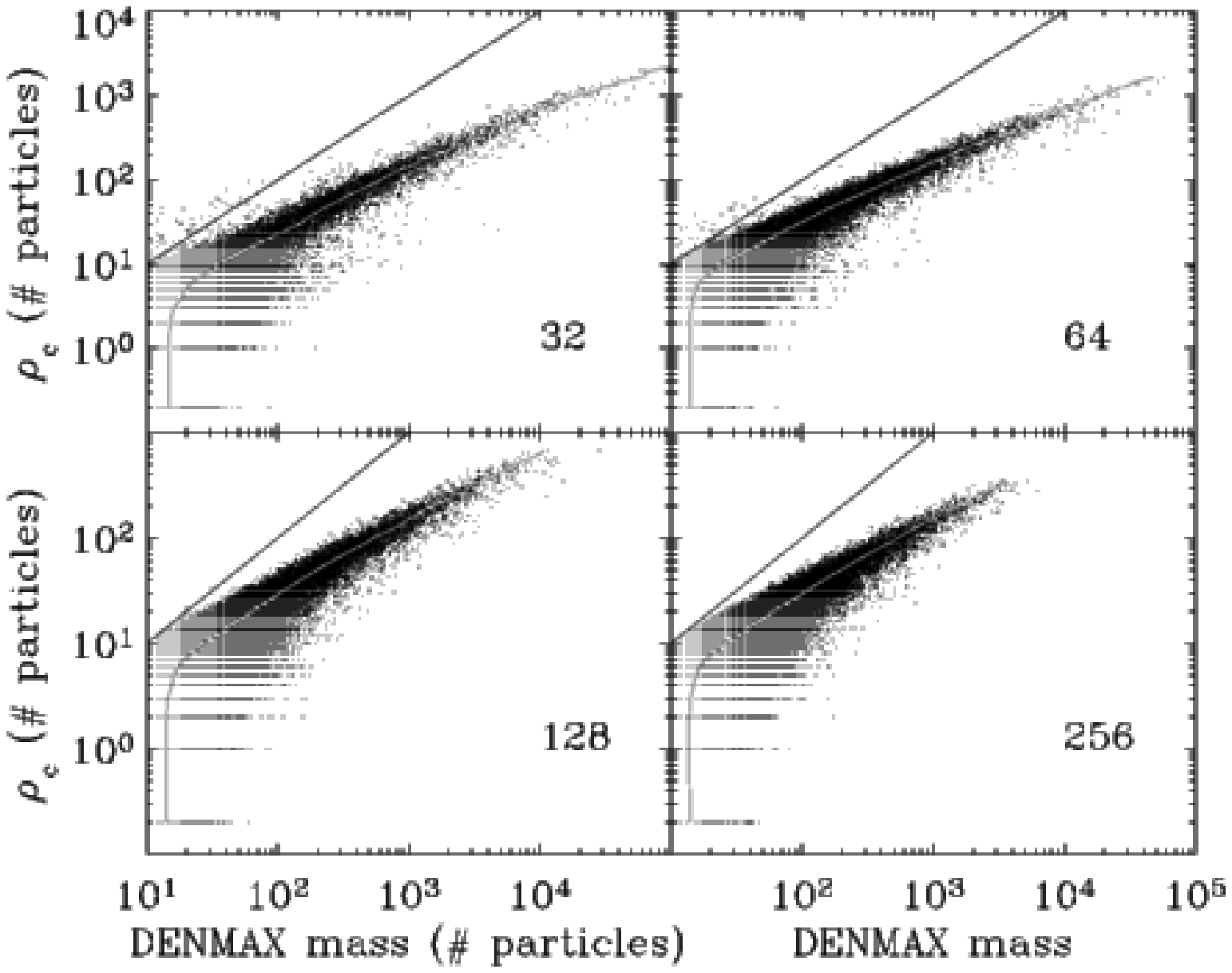}
    \end{center}
    \caption[1]{ \small Scatter plots of DENMAX mass (number of particles
      in the halo) versus central density $\rho_c$ (the number of particles
      within radius $r_\rho$) for all simulations.  Haloes with $\rho_c =
      0$ are shown with $\rho_c = 0.2$.  No haloes had a DENMAX mass under
      10 because DENMAX discards haloes with less than 10 particles.  The
      solid grey line passing through the distribution shows the average
      mass at each $\rho_c$.  To obtain the mass at a particular
      $\rho_{c,0}$, we averaged together the masses of haloes with $\rho_c$
      within $1/10$ of a dex of $\rho_{c,0}$.  The thin black line shows
      the identity function, $y=x$, which would result if all, and only,
      particles counted in the DENMAX mass were within $r_\rho$ of the halo
      centre.
    \label{mvrho}
    }
    \end{figure}
}

\newcommand{\vc}{
    \begin{figure}
    \begin{center}
    \leavevmode
    \epsfxsize=\columnwidth    
    \epsfbox{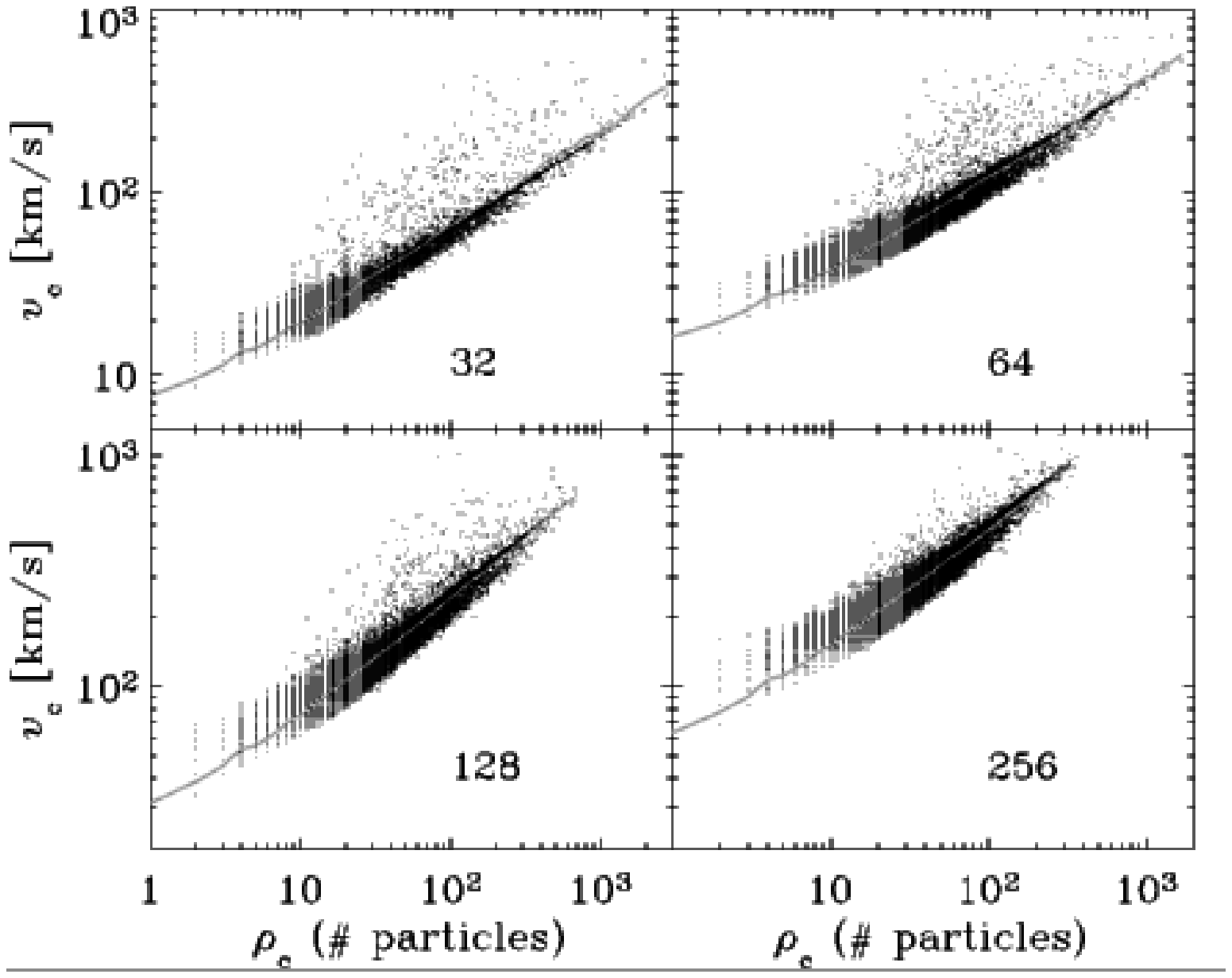}
    \end{center}
    \caption[1]{ \small Scatter plots of central density $\rho_c$ versus
      maximum circular velocity $v_c$ for all DENMAX haloes.  The solid grey
      line passing through the distribution shows the average $v_c$ at each
      $\rho_c$.  To obtain $v_c$ at a particular $\rho_{c,0}$, we averaged
      together $v_c$'s of haloes with $\rho_c$ within $1/10$ of a dex of
      $\rho_{c,0}$.
    \label{vc}
    }
    \end{figure}
}

\newcommand{\pwrfit}{
    \begin{figure}
    \begin{center}
    \leavevmode
    \epsfxsize=\columnwidth    
    \epsfbox{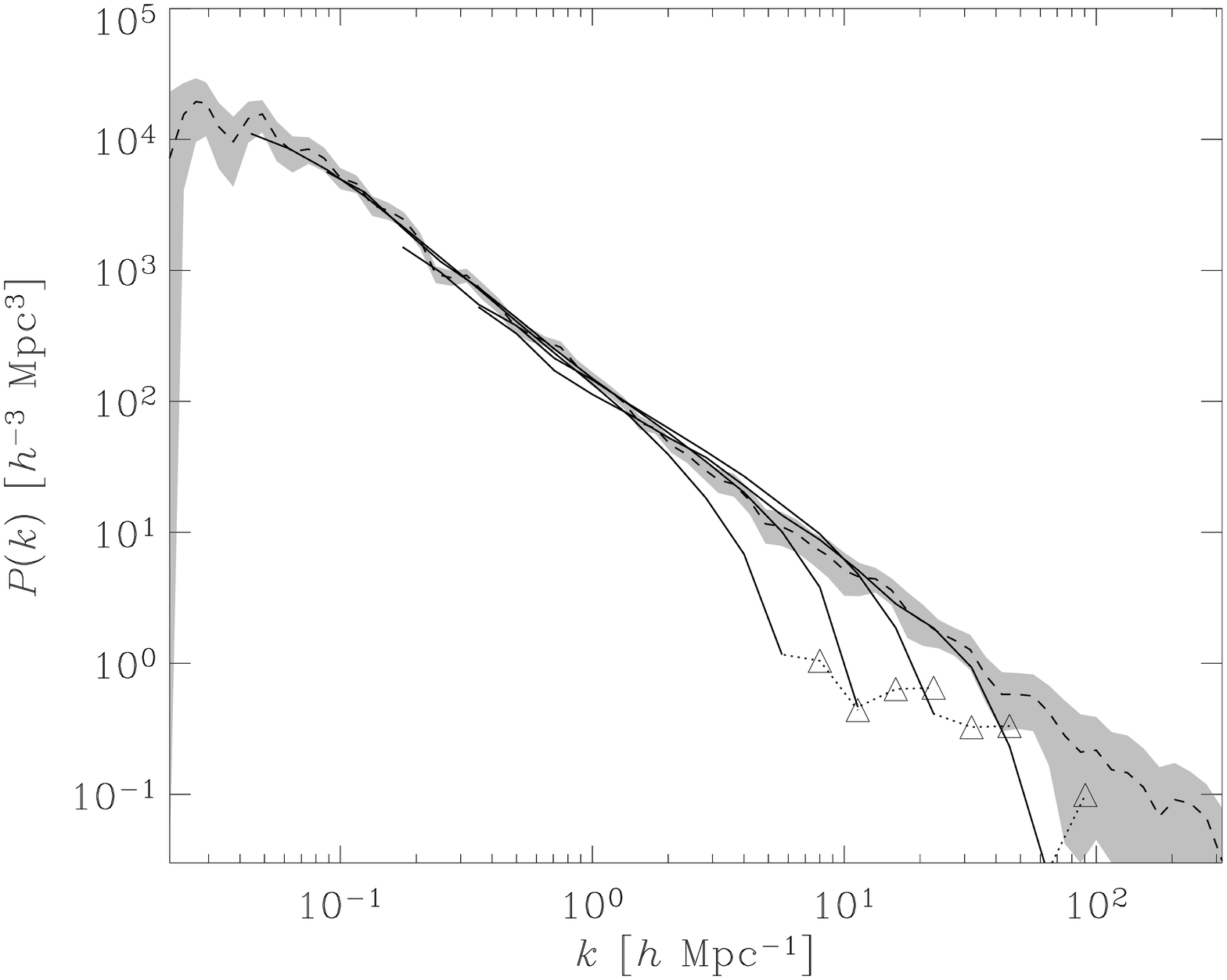}
    \end{center}
    \caption[1]{ \small The best-fitting power spectra (see
      Fig. \ref{individp} for error bars) for all simulations.  The PSC$z$
      power spectrum is the dashed curve with a grey error band; the best
      fits from the (from left to right) 256, 128, 64, and 32 $h^{-1}\!$
      Mpc simulations appear as solid curves.  Triangles connected with
      dotted lines denote negative values.
    \label{pwrfit}
    }
    \end{figure}
}
\newcommand{\pwrfittot}{
    \begin{figure}
    \begin{center}
    \leavevmode
    \epsfxsize=\columnwidth    
    \epsfbox{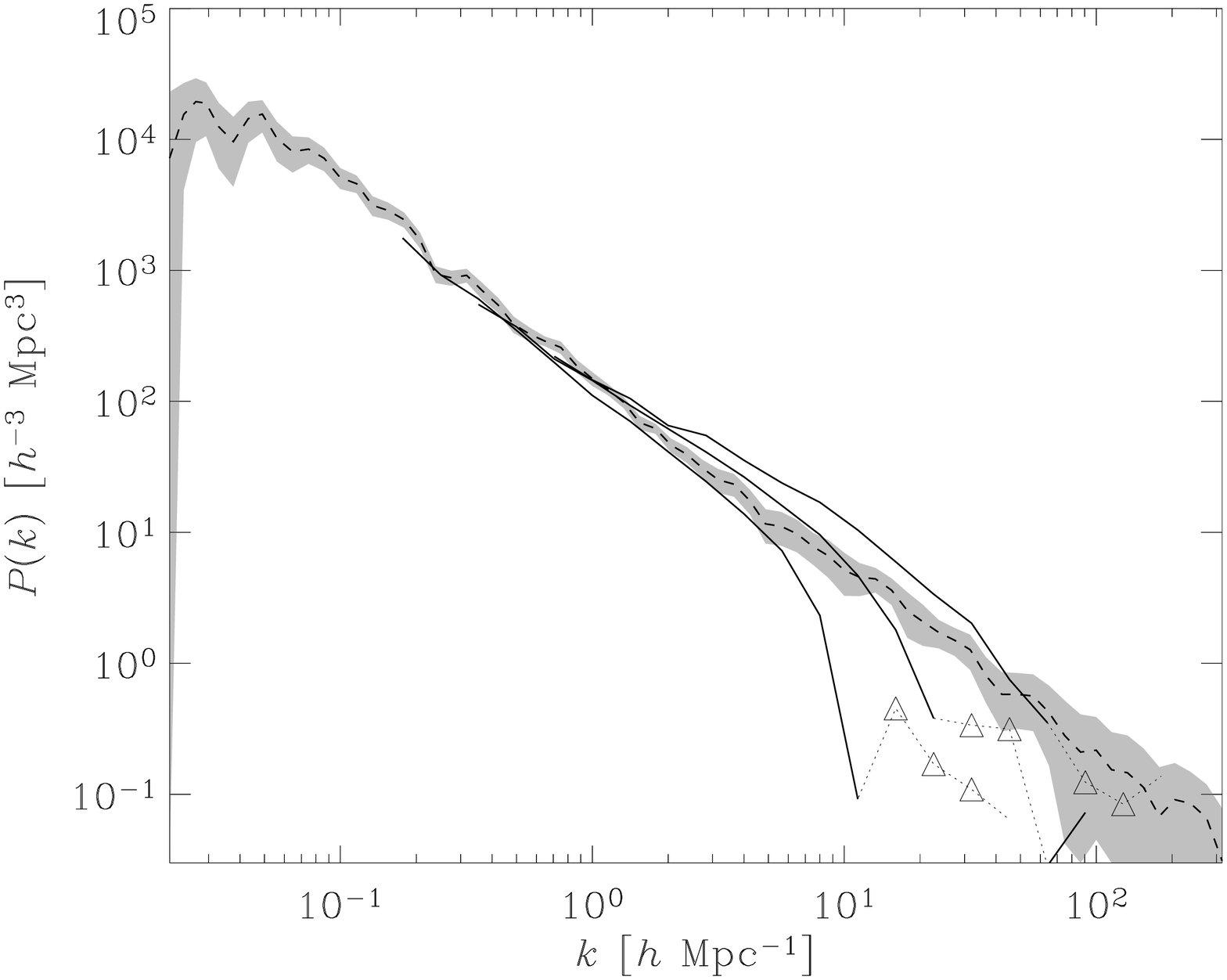}
    \end{center}
    \caption[1]{ \small The best-fitting power spectra (error bars are
      similar to those in Fig.\ \ref{individ}) for the 128, 64, and 32
      $h^{-1}\!$ Mpc simulations, constrained so that the halo number
      density matches.  The PSC$z$ power spectrum is the dashed curve with
      a grey error band.  Triangles connected with dotted lines denote
      negative values.
    \label{pwrfittot}
    }
    \end{figure}
}
\newcommand{\pseudochi}{
    \begin{figure}
    \begin{center}
    \leavevmode
    \epsfxsize=\columnwidth    
    \epsfbox{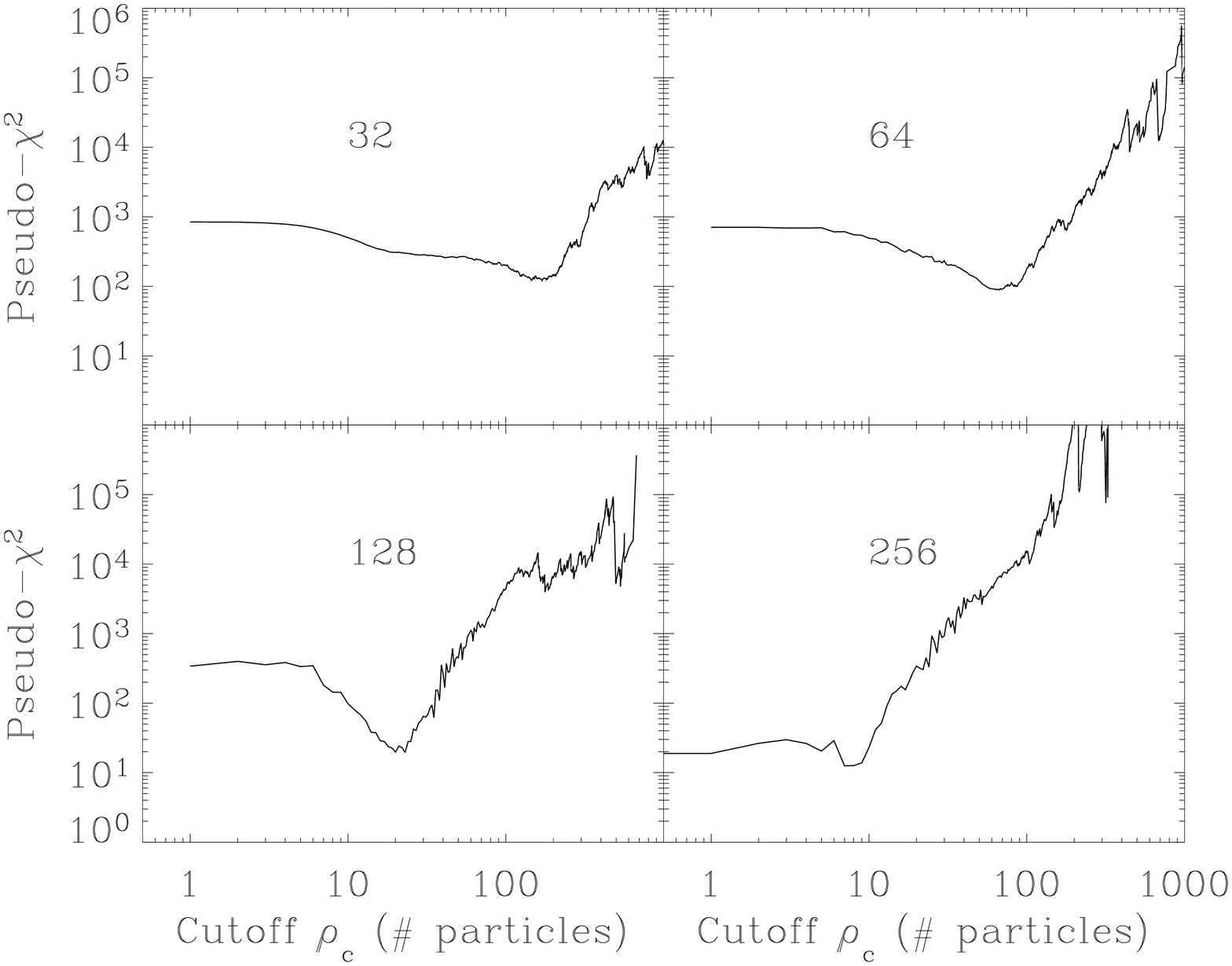}
    \end{center}
    \caption[1]{ \small $\tilde{\chi}^2$ values for the four simulations as
      functions of central density cutoff $\rho_{c,min}$.  In the 256
      $h^{-1}\!$ Mpc simulation, the $\tilde{\chi}^2$ value at
      $\rho_{c,min} = 0$ (including all detected haloes) is shifted to
      $\rho_{c,min} = 0.5$ so that it can appear on a log-log plot.
    \label{pseudochi}
    }
    \end{figure}
}
\newcommand{\pseudochin}{
    \begin{figure}
    \begin{center}
    \leavevmode
    \epsfxsize=\columnwidth    
    \epsfbox{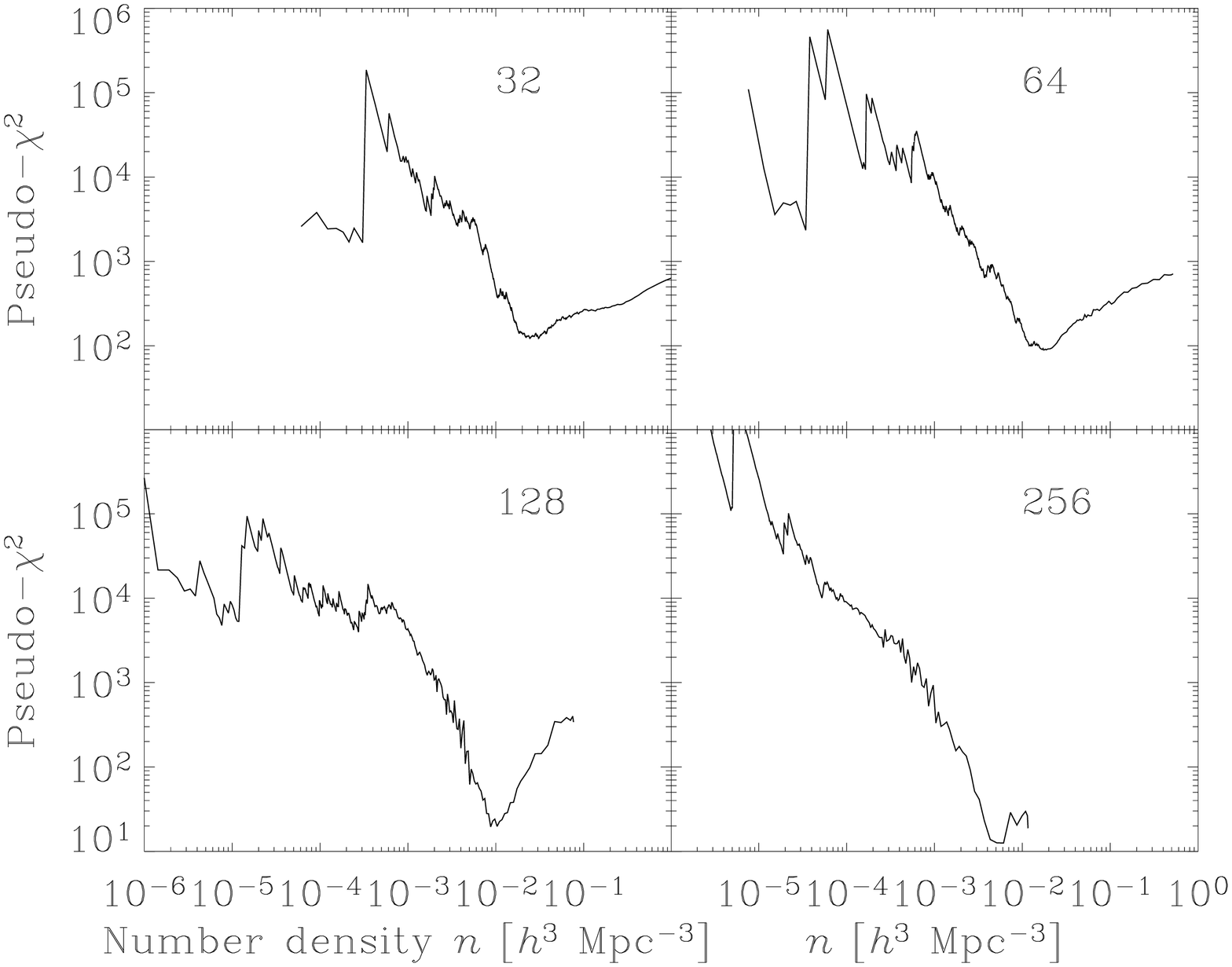}
    \end{center}
    \caption[1]{ \small $\tilde{\chi}^2$ values for the four simulations as
      functions of halo number density.  The best-fitting number densities
      are 0.0244, 0.0175, 0.00873, and 0.00611 haloes per $h^{-3}$ Mpc$^3$
      for the 32, 64, 128, and 256 $h^{-1}$ Mpc simulations, respectively.
    \label{pseudochin}
    }
    \end{figure}
}
\newcommand{\pseudochit}{
    \begin{figure}
    \begin{center}
    \leavevmode
    \epsfxsize=1.6in    
    \epsfbox{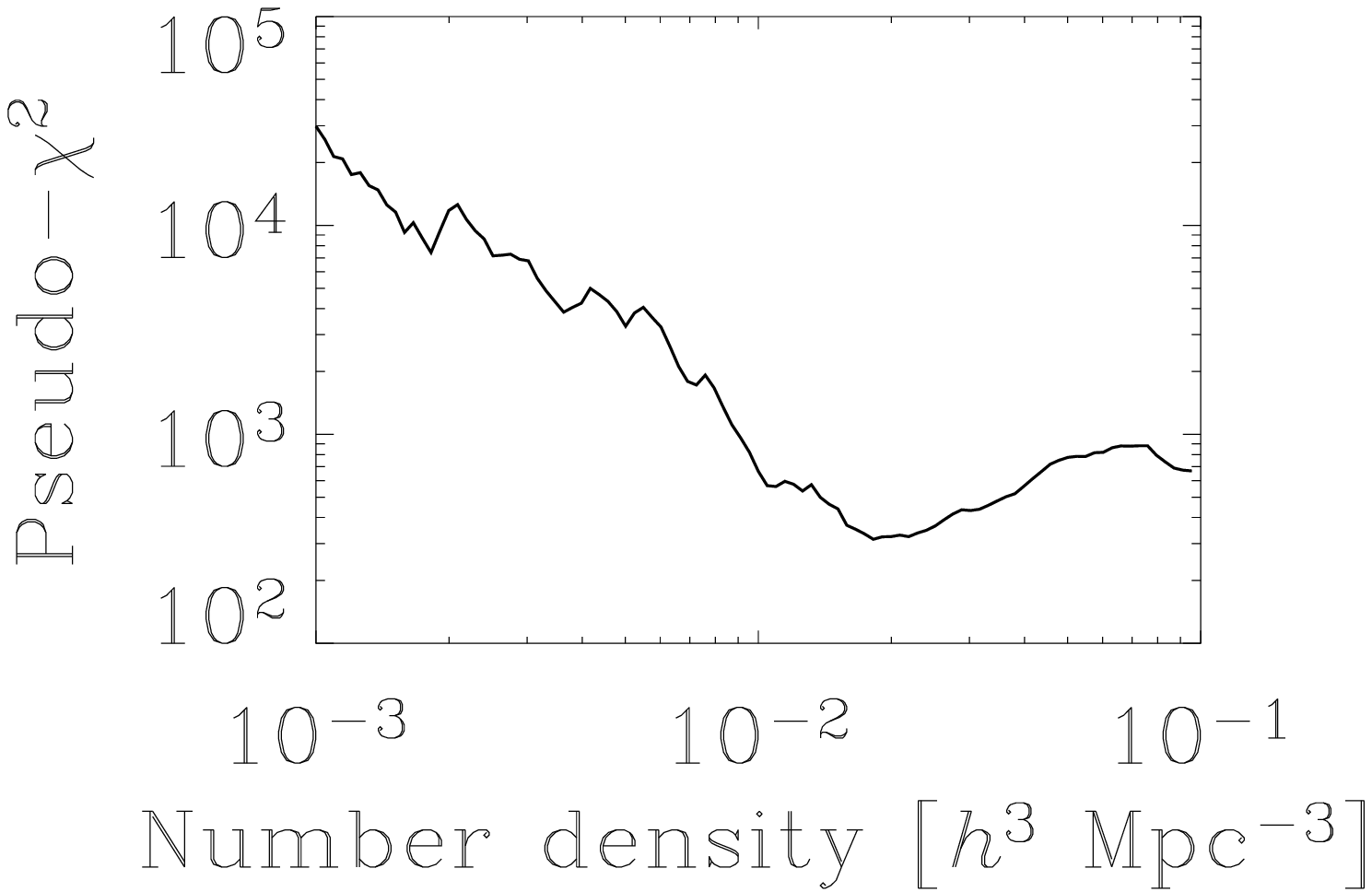}
    \end{center}
    \caption[1]{ \small The sum over the smallest three simulations of
      $\tilde{\chi}^2$ as a function of halo number density.  The best fit
      is at 0.0182 haloes per $h^{-3}$ Mpc$^3$, which corresponds to the
      PSC$z$ galaxy number density at a depth of about 20 $h^{-1}$ Mpc.
    \label{pseudochit}
    }
    \end{figure}
}
\newcommand{\individ}{
    \begin{figure}
    \begin{center}
    \leavevmode
    \epsfxsize=\columnwidth    
    \epsfbox{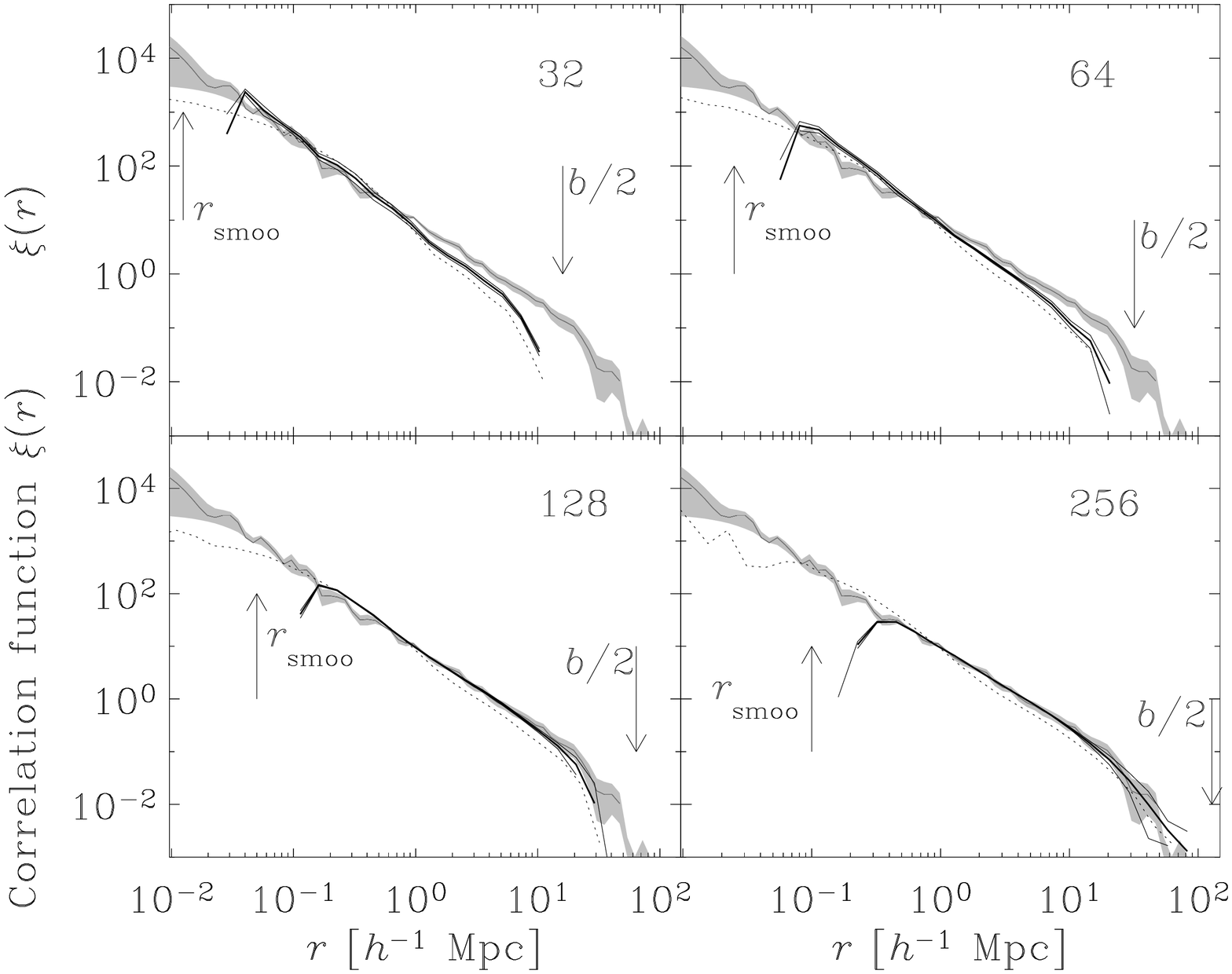}
    \end{center}
    \caption[1]{ \small The best-fitting halo correlation functions (thick
      solid curves) shown individually with error bands (thin solid
      curves).  The PSC$z$ correlation function appears as a light curve with
      grey error bands, and the simulations' dark matter correlation
      functions appear as dotted curves.  Arrows show the scales of the
      DENMAX$^2$ smoothing length $r_{\rm smoo}$ ($b/2560$), and half the
      box size ($b/2$), where the correlation function in a box of size $b$
      becomes meaningless.  All softening lengths were 0.01 $h^{-1}$ Mpc.
    \label{individ}
    }
    \end{figure}
}
\newcommand{\individp}{
    \begin{figure}
    \begin{center}
    \leavevmode
    \epsfxsize=\columnwidth    
    \epsfbox{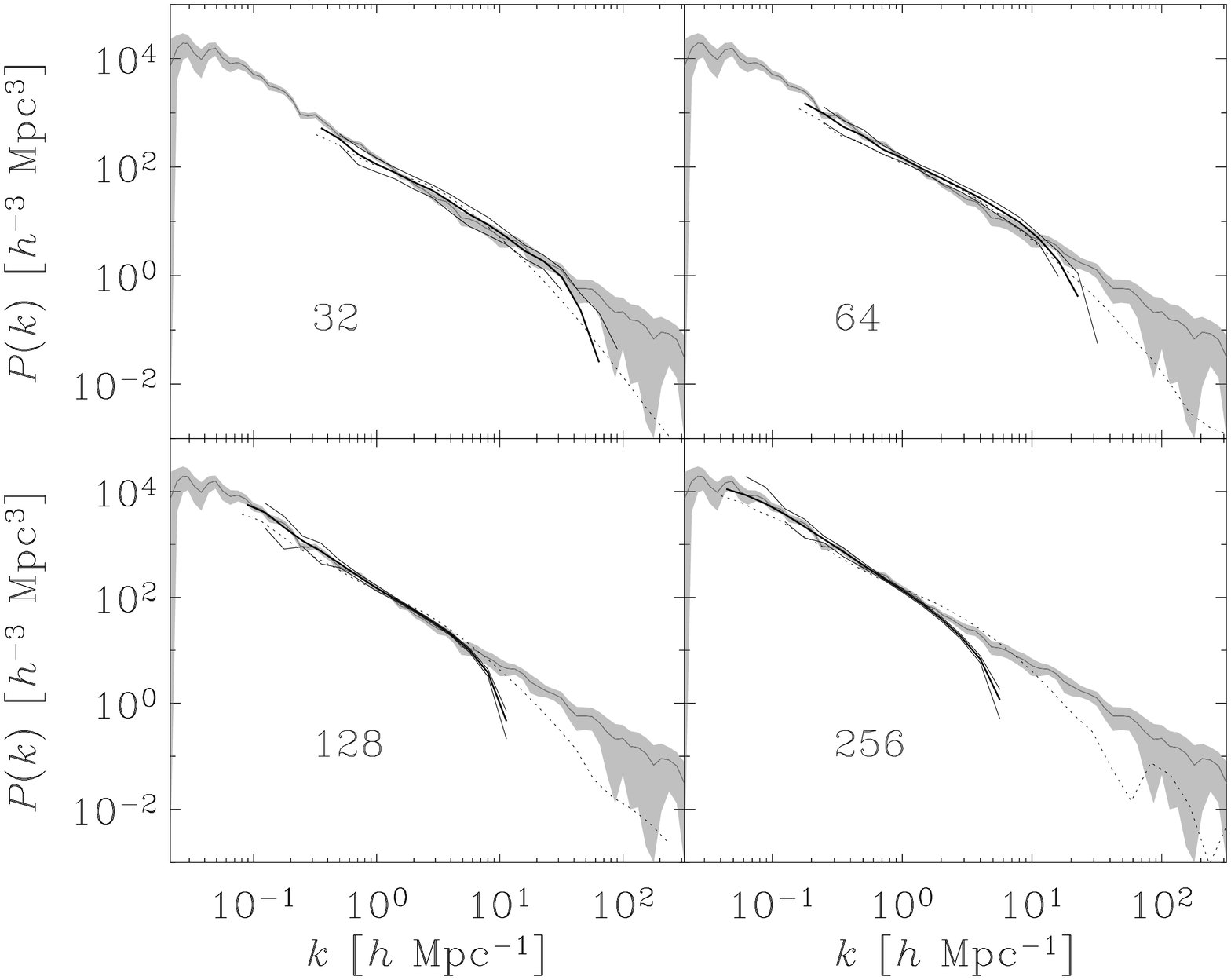}
    \end{center}
    \caption[1]{ \small The best-fitting halo power spectra (thick solid
      curves) shown individually with error bands (thin solid curves).  The
      PSC$z$ power spectrum appears as a light curve with grey error bands,
      and the simulations' dark matter power spectra appear as dotted
      curves.

      \label{individp}
    }
    \end{figure}
}
\newcommand{\bias}{
    \begin{figure}
    \begin{center}
    \leavevmode
    \epsfxsize=\columnwidth    
    \epsfbox{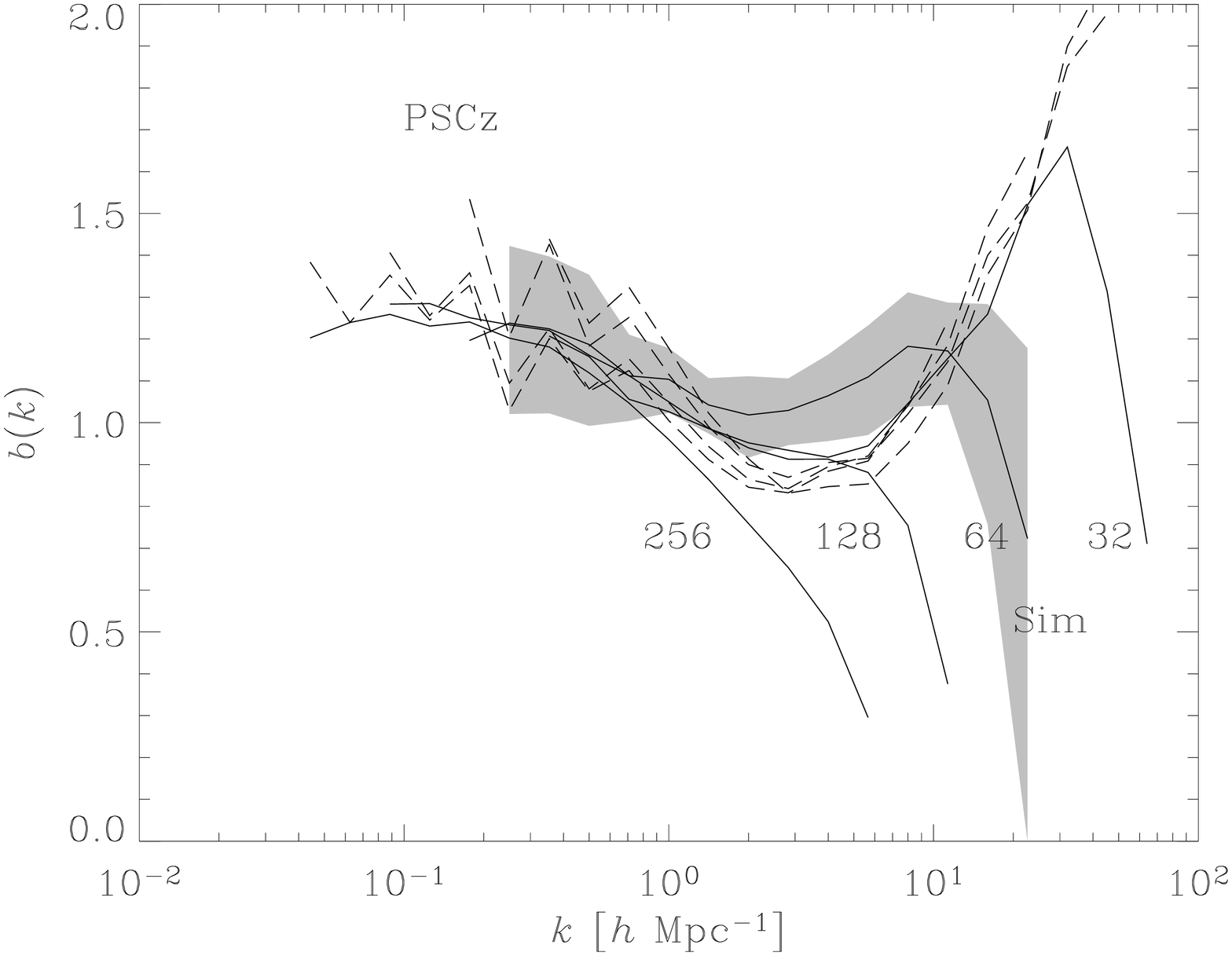}
    \end{center}
    \caption[1]{ \small The dashed curves show the bias factors between
      both the PSC$z$ power spectrum and the dark matter power spectra in
      each simulation; the solid curves show the bias factors between the
      haloes and the dark matter in each simulation.  The PSC$z$ bias
      curves are $b_{{\rm PSC}z}(k)=\sqrt{P_{{\rm PSC}z}(k)/P_{\rm
      dm}(k)}$, with a different $P_{\rm dm}$ for each simulation.  The
      simulation bias curves are $b_{\rm sim}(k)=\sqrt{P_{\rm
      haloes}(k)/P_{\rm dm}(k)}$, again running through the simulations.
      An error band arising from the errors in the power spectrum in Figure
      \ref{individp} as been put around the 64 $h^{-1}$ Mpc bias curve,
      representative of the others.
      \label{bias}
    }
    \end{figure}
}
\newcommand{\cenpic}{
    \begin{figure*}
    \begin{minipage}{175mm}
    \begin{center}
    \leavevmode
    \epsfxsize=\columnwidth   
    \epsfbox{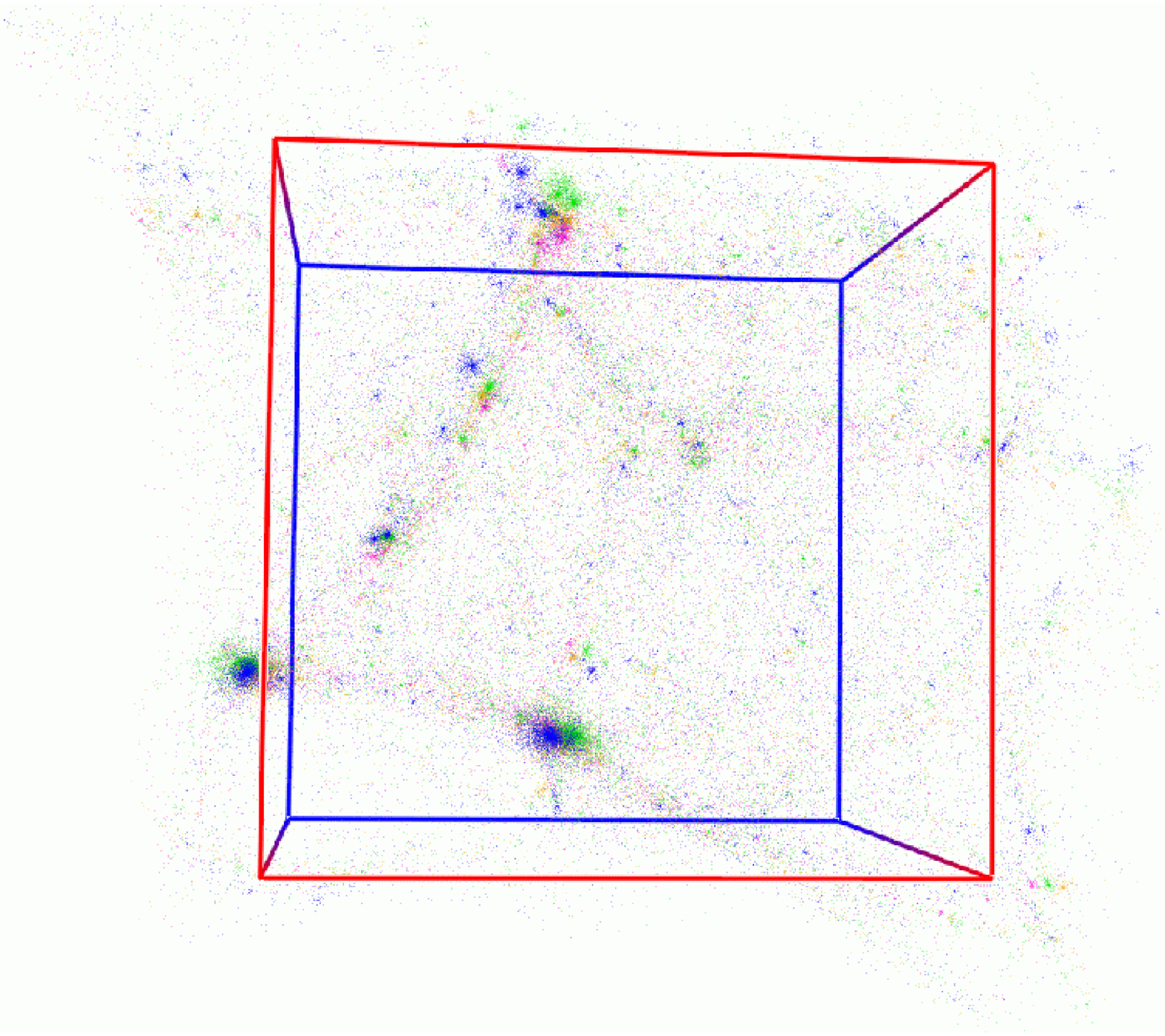}
    \end{center}
    \caption[1]{ \small The region originally in the central 16 $h^{-1}$
      Mpc region of each simulation.  Particles from the 32, 64, 128, and
      256 $h^{-1}$ Mpc simulations appear in magenta, orange, green, and
      blue, respectively.  The sets of particles from simulations of box
      size greater than 32 $h^{-1}$ Mpc have undergone a translation and
      linear transformation to lie on the set from the (untransformed) 32
      $h^{-1}$ Mpc simulation.  All particles from the central region are
      shown for the 128 $h^{-1}$ Mpc simulation, and the number densities
      of particles from other simulations are adjusted by factors of eight
      to match.  Thus, 1/8 and 1/64 of the particles from the 64 and 32
      $h^{-1}$ Mpc simulations appear.  In the 256 $h^{-1}$ Mpc case, the
      number density was octupled by adding new particles which bisect
      lines joining particles adjacent on the initial mesh.  The
      predominantly blue character of the large haloes at the bottom arises
      because the halo in the 256 $h^{-1}$ Mpc simulation is simply in
      front of the others.  The figure was produced using Nick Gnedin's
      {\tt IFRIT} visualization tool, at
      \url{http://casa.colorado.edu/~gnedin/IFRIT/}.
      \label{cenpic}
    }
    \end{minipage}
    \end{figure*}
}
\newcommand{\pd}{
    \begin{figure}
    \begin{center}
    \leavevmode
    \epsfxsize=\columnwidth    
    \epsfbox{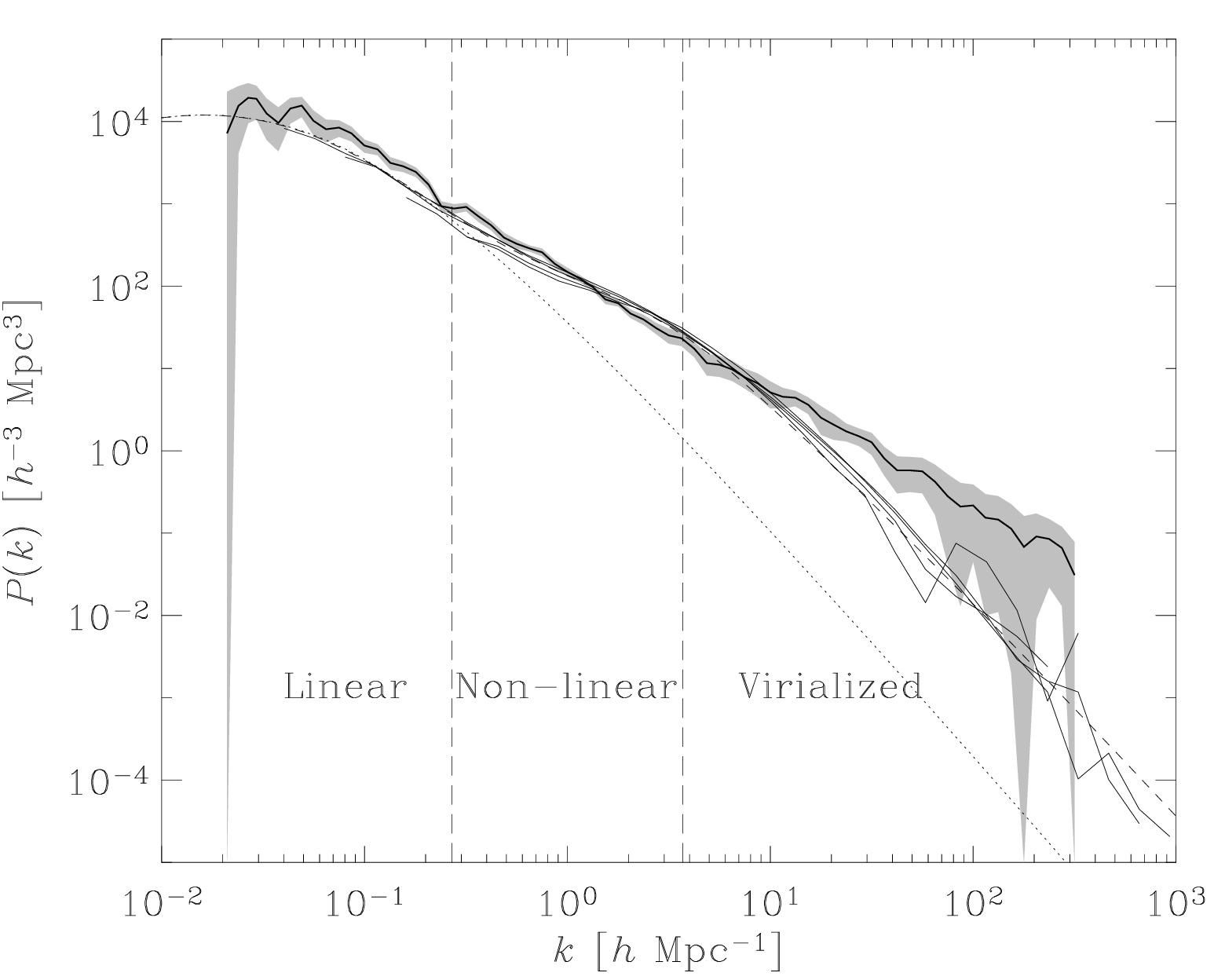}
    \end{center}
    \caption[1]{ \small Comparison of the dark matter power spectra from
      the simulations (thin solid curves) to the linear power spectrum
      (dotted curve) and, for the sake of illustration, an estimate of the
      non-linear power spectrum (dashed curve) evolved from the linear
      power spectrum using the method of Peacock \& Dodds (1996).  The
      PSC$z$ power spectrum also appears, surrounded by a grey band.  The
      boundaries of the `non-linear' range of wavenumber were placed by
      hand, approximately at the two inflections in the dark-matter power
      spectrum.
      \label{pd}
    }
    \end{figure}
}
\newcommand{\halofig}{
    \begin{figure*}
    \begin{minipage}{175mm}
    \begin{center}
    \leavevmode
    \epsfxsize=\columnwidth    
    \epsfbox{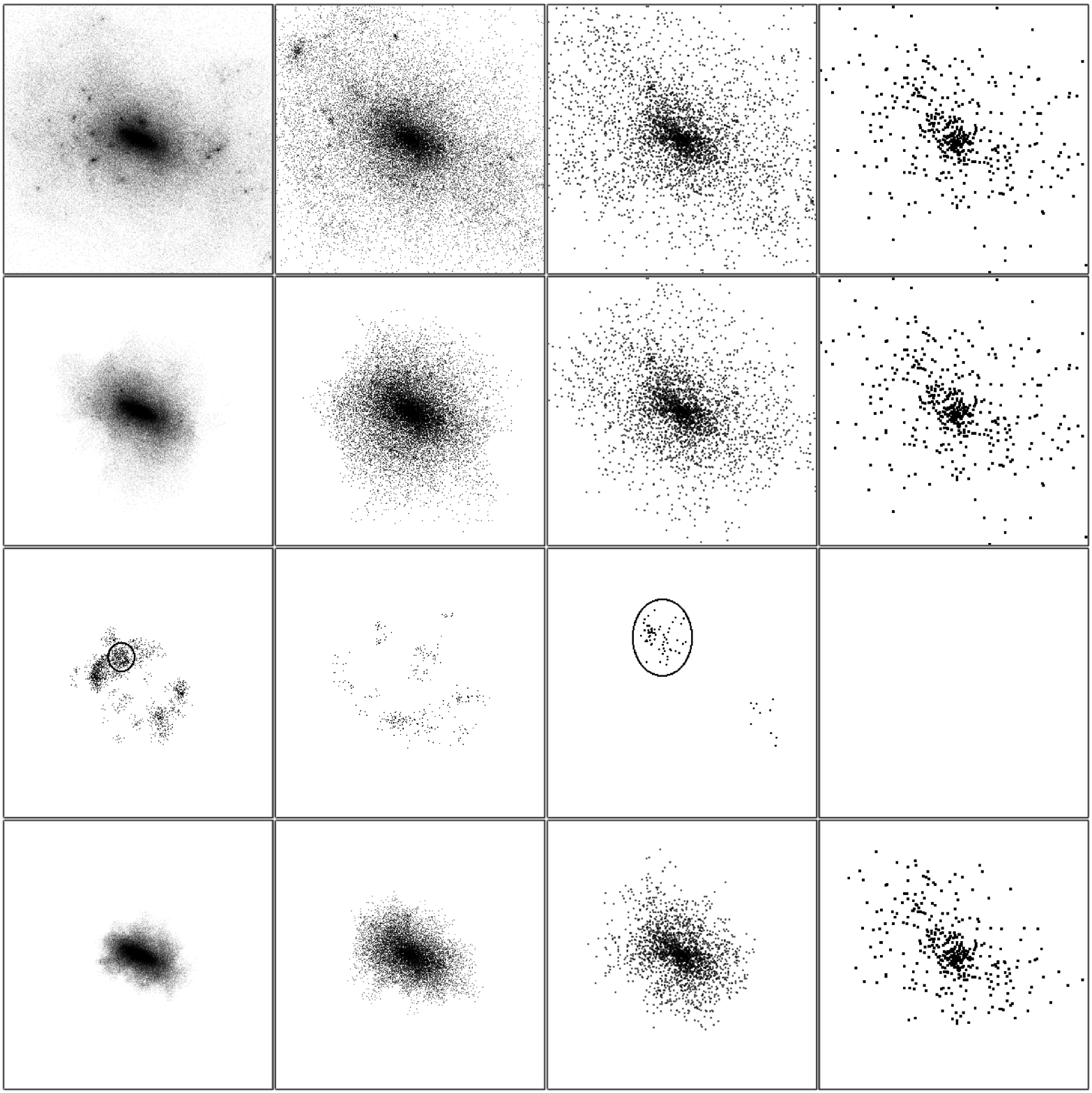}
    \end{center}
    \caption[1]{ \small Comparison of a halo (the third-largest in the 32
      $h^{-1}$ Mpc simulation, with a mass of a few times 10$^{13}$
      M$_\odot$) in different simulations, and the effect of DENMAX$^2$ on
      it.  The first row shows all particles, whether they were detected by
      DENMAX or not, in a box 1.8 $h^{-1}$ Mpc on a side around the center
      of the halo, in, from left to right, the 32, 64, 128, and 256
      $h^{-1}$ Mpc simulations.  The second row shows particles in the halo
      as detected by DENMAX.  Note that there are many evident `subhaloes'
      which do not appear in the second row; these were already detected as
      different individual haloes by DENMAX, before application of
      DENMAX$^2$.  DENMAX$^2$ split the DENMAX halo into 55, 20, 3, and 1
      subhalo(es) in the 32, 64, 128, and 256 $h^{-1}$ Mpc simulations,
      respectively.  The satellite subhaloes appear in the third row, with
      circles around the subhaloes (one in each of the 128 and 32 $h^{-1}$
      Mpc simulations) satisfying the central density cutoff.  The final
      row shows the main subhalo (also satisfying the central density
      cutoff) as detected by DENMAX$^2$.  Note the physical enlargement of
      the halo with boxsize; this is a result of the increase in the
      smoothing length in physical units.  The figure was produced using
      Nick Gnedin's {\tt IFRIT} visualization tool, at
      \url{http://casa.colorado.edu/~gnedin/IFRIT/}.
      \label{halofig}
    }
    \end{minipage}
    \end{figure*}
}

\newcommand{\halocmp}{
    \begin{figure}
    \begin{center}
    \leavevmode
    \epsfxsize=\columnwidth    
    \epsfbox{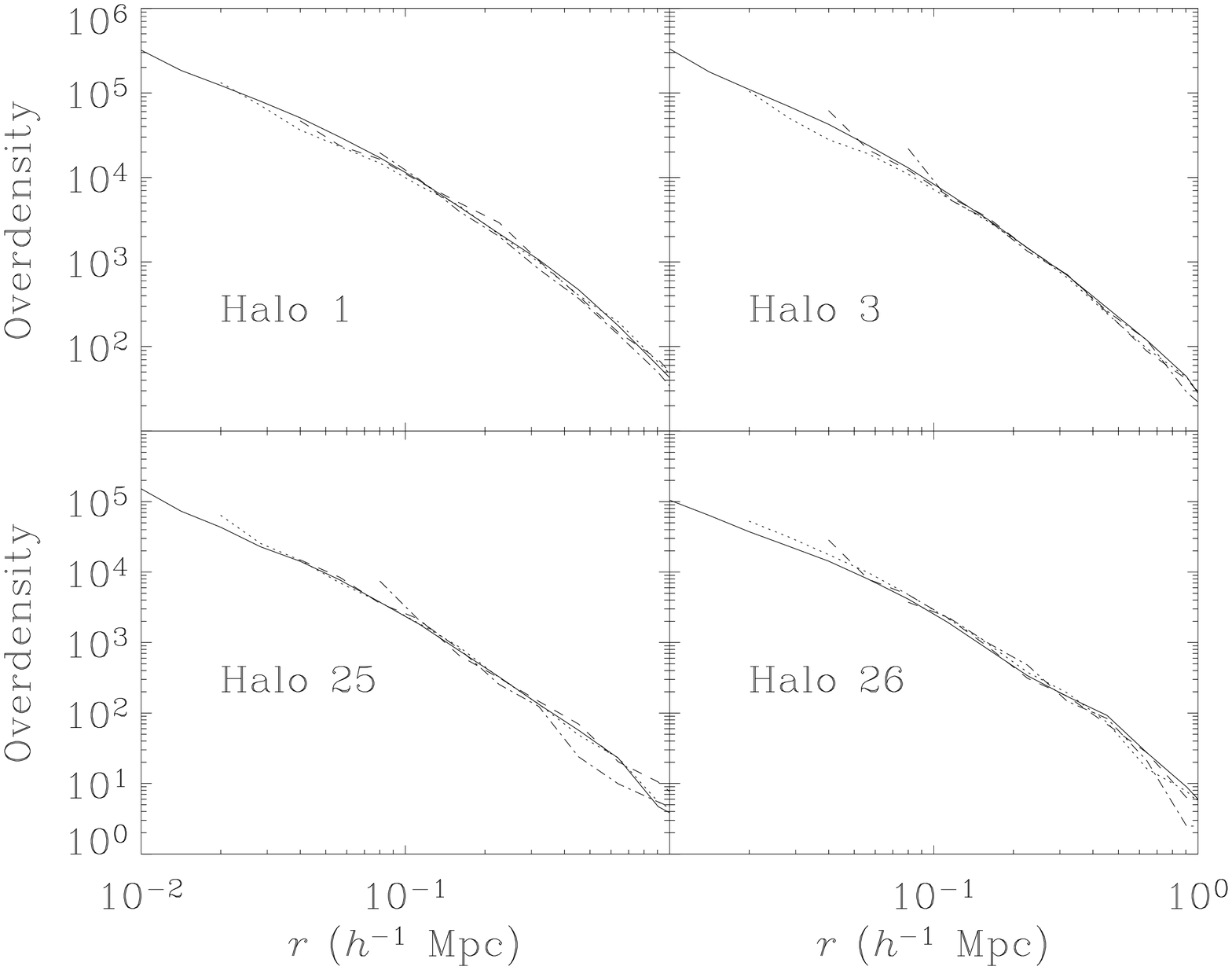}
    \end{center}
    \caption[1]{ \small Density profiles, as a function of distance from
    their DENMAX centres, of the four haloes identified across the
    simulations.  The solid, dotted, dashed, and dashdot curves are from the
    32, 64, 128, and 256 $h^{-1}\!$ Mpc simulations, respectively.  The
    halo appearing in Fig.\ \ref{halofig} is in the bottom-right corner.
    \label{halocmp}
    }
    \end{figure}
}

\newcommand{\cencrl}{
    \begin{figure}
    \begin{center}
    \leavevmode
    \epsfxsize=\columnwidth    
    \epsfbox{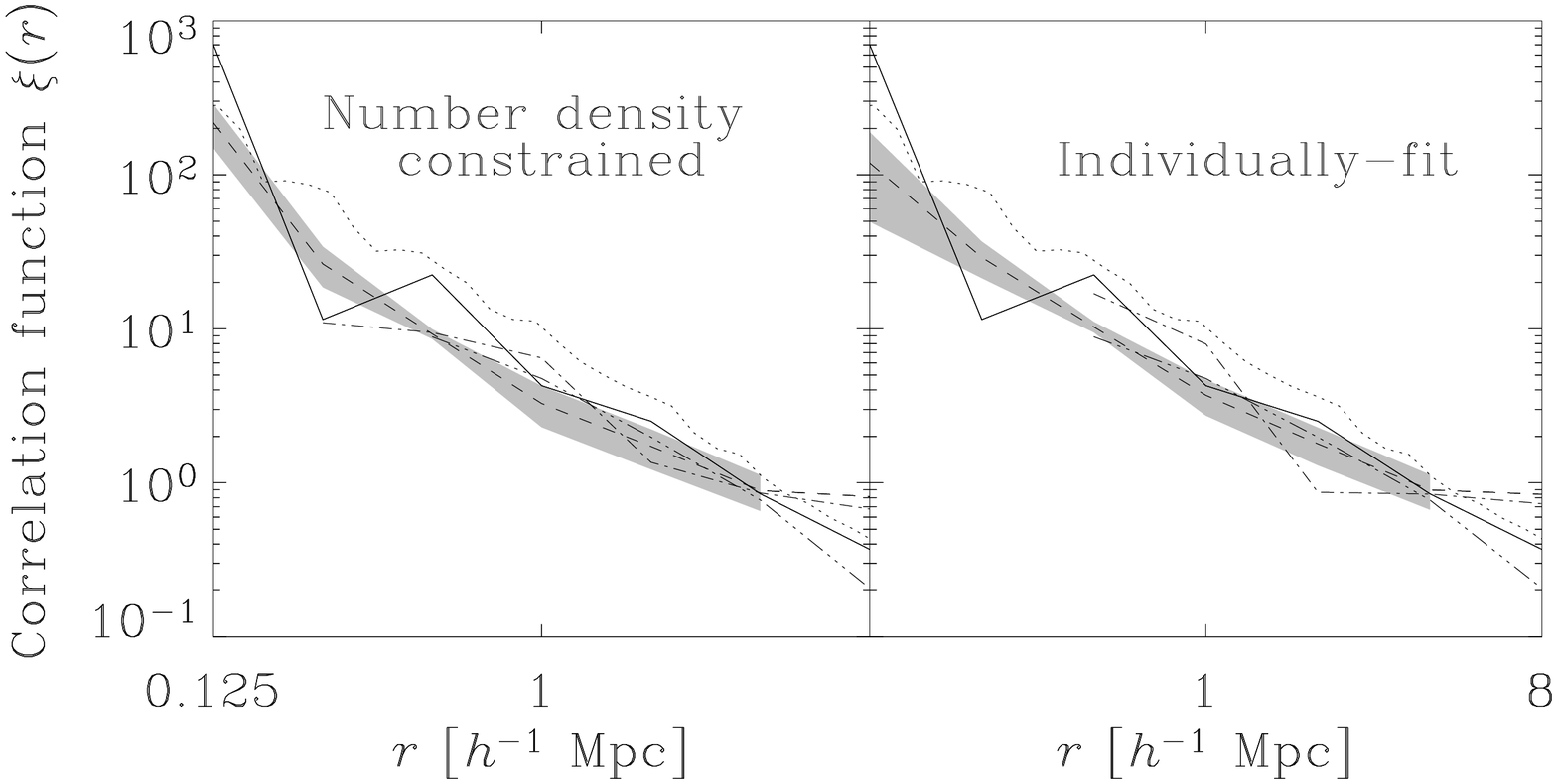}
    \end{center}
    \caption[1]{ \small Best-fitting halo correlation functions measured from
    the central cube of length 16 $h^{-1}$ Mpc from each simulation, when
    constrained to have matching number densities and when individually
    fit.  The larger halo sets have been mapped using the shift and
    deformation described at the end of \S \ref{sim} on to the 32 $h^{-1}$
    Mpc central region before finding the correlation function.  To boost
    the number of pairs, we actually calculated the cross-correlation of
    haloes in the central boxes with all haloes in the box in each case.
    The PSC$z$ correlation function is the dotted curve, and correlation
    functions from the 32, 64, 128, and 256 $h^{-1}$ Mpc simulations appear
    as solid, dashed, dashdotted, and dashdotdotdotted curves,
    respectively.  We have put error bands (measured as previously by
    dividing the region into octants) around the 64 $h^{-1}$ Mpc
    correlation function, which are typical of the others.
    \label{cencrl}
    }
    \end{figure}
}

\title[The PSC$z$ Galaxy Power Spectrum from N-body
Simulations]{Understanding the PSC{\mbox{\boldmath$z$}} Galaxy Power Spectrum with N-body
Simulations}

\author[Mark C.\ Neyrinck, Andrew J.\ S.\ Hamilton, and Nickolay Y.\ Gnedin]
{Mark C. Neyrinck$^{1,3}$, Andrew J.S. Hamilton$^{1,3}$, and Nickolay Y. Gnedin$^{2,3}$ \\
$^{1}$JILA, University of Colorado, Boulder, CO 80309\\
$^{2}$Center for Astrophysics and Space Astronomy, University of Colorado,
  Boulder, CO 80309\\
$^{3}$Department of Astrophysical and Planetary Sciences, University of
  Colorado, Boulder, CO 80309 \\
email: {\tt Mark.Neyrinck@colorado.edu}}
\begin{document}
\date{2003 August 28}

\maketitle

\begin{abstract}
  By comparing the PSC$z$ galaxy power spectrum with the results of nested
pure dark-matter N-body simulations, we try to understand how
infrared-selected galaxies populate dark-matter haloes, paying special
attention to the method of halo identification in the simulations.  We thus
test the hypothesis that baryonic physics negligibly affects the
distribution of galaxies down to the smallest scales yet observed.  We are
successful in reproducing the PSC$z$ power spectrum on scales $\la 40$ $h$
Mpc$^{-1}$, near our resolution limit, by imposing a central density cutoff
on simulated haloes, which gives a rough minimum mass and circular velocity
of haloes in which PSC$z$ galaxies formed.
\end{abstract}

\begin {keywords}
  large-scale structure of the universe -- galaxies: haloes --
  cosmology: theory -- galaxies: infrared -- galaxies: formation --
  methods: {\it N}-body simulations.
\end {keywords}
    
\section{Introduction}

The distribution of ice in the ocean can be difficult to study when only
the tips of icebergs can be observed.  We can catalogue the positions and
redshifts of galaxies, and can obtain glimpses of the intergalactic
environment by observing the Lyman-alpha forest, but the dark matter, the
component which plays the largest role in our current paradigm of structure
formation, remains obscure.  In this paper, we try to connect underlying
dark-matter ice to the iceberg tips (galaxies) we can see.

Only in the last few years have we claimed to have a successful
cosmological model, the `concordance' $\Lambda$CDM model.  Its loose ends
do seem tieable, but some areas remain largely mysterious: for example, the
formation of galaxies within dark-matter haloes, and the resulting
relationship between them at the present epoch.  Here, we investigate this
relationship through their distributions, most succinctly quantified by the
power spectrum, or its Fourier dual, the correlation function.  Currently,
the most extensive measurement of a power spectrum of observed galaxies,
ranging over 4.5 decades of wavenumber, is by Hamilton \& Tegmark (2002,
hereafter HT).  It was made from the PSC$z$ (Point Source Catalogue
Redshift) catalogue (Saunders et al.\ 2000) of galaxies observed with the
{\it IRAS} infrared satellite.  There will soon be a flood of galaxy
clustering data, for example from the Sloan Digital Sky Survey (SDSS, York et
al.\ 2000) and the Two-Degree Field (Lewis et al.\ 2002) survey.  Early
data (e.g.\ Zehavi et al.\ 2002, Percival et al.\ 2001) suggest that
clustering properties vary with galaxy morphology, luminosity, and colour.
Here we restrict ourselves to PSC$z$ infrared galaxies, but with excellent
optical data differentiated by colour, an approach such as ours will soon be
able to say more about the types of galaxies which inhabit different sorts
of dark-matter haloes.

As with other measurements of galaxy power spectra, for example from the
APM galaxy survey (Baugh 1996), HT found a roughly power law form.  This is
strikingly different from the dark matter power spectrum in the current
concordance $\Lambda$CDM cosmological model.  Figure \ref{pd} shows the
PSC$z$ power spectrum along with linear and non-linear power spectra for
the concordance $\Lambda$CDM model, and also dark matter power spectra from
our simulations.  The non-linear dark matter spectrum traces the linear
spectrum at large scales, but at smaller, non-linear scales, it rises above
it because waves on the scale of collapsing structures grow faster than
waves in the linear regime.  At even smaller scales, virialization slows
growth, producing a downward inflection.

\pd

The galaxy and dark-matter power spectra thus appear to be biased with
respect to each other (i.e.\ they are different); their different shapes
indicate that bias is scale-dependent.  Numerous attempts have been made to
understand this.  The halo model of large-scale structure (reviewed by
Cooray \& Sheth, 2002) assumes the existence of small, bound objects
(haloes) which are clustered according to the linear power spectrum.
Galaxy clustering statistics such as the correlation function can then be
calculated as the sum of two terms describing pairs of galaxies from the
same and from different haloes.  This can make use of a Halo Occupation
Distribution, HOD (Benson 2001; White, Hernquist \& Springel 2001; Berlind
\& Weinberg 2002; Berlind et al.\ 2002), which describes the number of
galaxies which inhabit a halo of a given mass.  In this context, a halo is
defined as a region at least 200 (typical of virialization) times more
dense than the background.  The approximations of the halo model permit
analytic models for the bias between galaxies and dark matter (e.g.\ Seljak
2000; Sheth \& Jain 2002).  Recently, a slight but statistically
significant deviation from a power law in the projected correlation
function of a preliminary SDSS sample was successfully modelled using a HOD
in the halo model (Zehavi et al.\ 2003).

Semi-analytic galaxy formation models (e.g.\ White \& Frenk 1991; Kauffmann
et al.\ 1999; Benson et al.\ 2000; Somerville et al.\ 2001, Mathis et al.\
2002) and hydrodynamic simulations (e.g.\ Katz, Hernquist \& Weinberg 1992;
Cen \& Ostriker 2000; Dav\'{e} et al.\ 2000; Pearce et al.\ 2001; White,
Hernquist, \& Springel 2001; Yoshikawa et al.\ 2001) have also been
successfully applied to the problem of bias.  While these approaches
attempt to give the relationship between galaxies and dark matter directly,
some of the galaxy formation prescriptions in semi-analytic models can be
rather ad hoc, and it is not clear that hydrodynamic simulations correctly
treat every piece of relevant baryonic physics.

In this paper, we take a different tack: we directly fit the PSC$z$ power
spectrum to dark-matter-only N-body simulations.  We are thus seeing how
far we can take the assumption that on intergalactic scales, baryonic
physics negligibly affects the clustering of haloes which contain observed
galaxies.  Similar studies, without fitting to specific observations, and
using different halo-finding algorithms, have been undertaken by Kravtsov
\& Klypin (1999) and Col\'{\i}n et al.\ (1999).

We also do not employ a HOD framework.  Such a statistical placement of
galaxies in haloes is useful in constructing an analytic model or when
faced with poor resolution, and can aid intuition.  Our approach is more
direct: if a halo is defined not as a region above a certain overdensity,
but as a region gravitationally bound to a significant maximum in the
density field, a definition which would admit subhaloes without their
parent haloes, then the number of galaxies detected by a redshift survey
inside a halo is either zero or one.

We pay particular attention to the halo-finding algorithm we use to go from
a dark matter distribution to a distribution of haloes, and to the effects
of resolution on the algorithm.  This is the most nontrivial step in
comparing the results of simulations to the observed galaxy power spectrum,
so it is important to be careful.

\section{Method}
\subsection{The Simulations}\label{sim}
The PSC$z$ power spectrum spans $4.5$ decades of wavenumber; to replicate
this dynamic range in an N-body simulation with sufficient mass resolution
would be unfeasible.  For this reason, and also to test for resolution
effects, we ran four manageably sized 256$^3$-particle simulations, of
comoving box size 32, 64, 128, and 256 $h^{-1}\!$ Mpc, with an adaptive
P$^3$M code (Bertschinger 1991).  A simulation of box size less than about
32 $h^{-1}\!$ Mpc would miss significant tidal forces from large-scale
fluctuations, and also could not form large clusters that appear with low
number densities in nature.  The values of cosmological parameters we used
in the simulation were from the concordance model of Wang, Tegmark \&
Zaldarriaga (2002): $\Omega_m = 0.34$, $\Omega_\Lambda = 0.66$, $h = 0.64$,
and $n = 0.93$.  We calculated the transfer function of the initial
conditions with the code of Eisenstein \& Hu (1999), which returned a value
of $\sigma_8 = 0.63$, the rms fluctuation of mass in spheres of radius 8
$h^{-1}\!$ Mpc.

There were two resolution issues to consider: mass resolution and spatial
resolution.  The mass resolution, i.e.\ the mass of a particle in the
simulation, depends on the number of particles per unit volume.  (See Table
\ref{fits} for particle masses for each simulation.)  So, the mass
resolution in the four simulations necessarily changes with box size.
However, we decided to use the same spatial resolution (softening length)
for all simulations: 10 $h^{-1}\!$ kpc, roughly the lowest scale probed in
the PSC$z$ power spectrum.

\begin{table*}
  \caption{ \small Choice of initial conditions.  Both the PSC$z$ galaxy
    density and the `Simulation' dark matter density are normalized to
    one.}
  \label{init}
  \begin{tabular}{lllll}
               & PSC$z$ & Simulation & PSC$z$ & Simulation\\
  Corner boxes & 1.49255 & 1.11328 & 0.83325 & 0.50146\\
               & 0.49410 & 0.76367 & 0.31982 & 0.47900\\
               & 1.45612 & 1.20386 & 2.35201 & 1.99805\\
               & 1.60033 & 1.63550 & 0.39196 & 0.83447\\
  Central box  & 1.36050 & 1.10596\\
\end{tabular}
\end{table*}

Unlike many cosmological simulations, our softening length was fixed in
physical, not comoving (Eulerian, not Lagrangian), coordinates.  This means
that at early epochs, the comoving softening length was larger, at maximum
about $1/6$ the mean interparticle separation in the 32 $h^{-1}\!$ Mpc
simulation, and less by factors of two in the others.  The first haloes to
collapse at $z\approx 10$ turn out to be quite important in determining the
fine structure of haloes at $z=0$.  The first haloes contain only a few
particles at $z \approx 10$, and a small comoving smoothing length can make
their relaxation times tiny, resulting in an overproduction of small,
relaxed structures (Moore 2001; Binney \& Knebe 2002).  By having a fairly
large comoving softening length at early times, we hope to have mitigated
this problem.

We wanted the four simulated regions to be as similar as possible under the
constraints of periodic boundary conditions, enabling us to compare
structures in the four simulations to each other.  Bertschinger (2001),
building on the work of Pen (1997), developed a simple formalism to
generate Gaussian random fields with multiple levels of resolution, which
we used to nest the initial conditions.  Effectively, this means that the
phases of fluctuations matched in the centres of each set of initial
conditions.  An animation depicting the nesting of the boxes can be found
at \url{http://casa.colorado.edu/~neyrinck/nesthalf.mpg}.  The
zone of agreement between two simulations of box size $b$ and $b/2$ is a
central cube of side length roughly $b/4$; a greater zone of agreement is
not possible since the smaller box has to obey periodic boundary
conditions.

We also selected our initial conditions so that the central region common
to the four simulations would have a structure similar to the Local Group
and its environs.  In this way, we hoped to replicate some features of the
way the PSC$z$ observations were made, looking out from the Milky Way, in a
slightly overdense region on the outskirts of a modest-sized supercluster.
The highest-resolution information in the simulations is from the smallest
box, just as the galaxies closest to each other in the PSC$z$ catalogue came
from regions close to the Milky Way.

To obtain these special initial conditions, we generated ten sets of
initial conditions for a 256 $h^{-1}\!$ Mpc simulation on a 256$^3$ mesh,
and ran them to the present epoch using a (fast but low-resolution) PM
code.  We counted particles in each cell of side length 8 $h^{-1}$ Mpc, the
traditional scale of non-linearity, over which we can trust the results of
the quick PM code.  This gave a 32$^3$ grid of dark matter density
estimates $\rho$ at the present epoch from each set of initial conditions.
The same was done for the PSC$z$ catalogue, binning galaxies on a $4^3$ grid
of 8 $h^{-1}$ Mpc cells with the Milky Way at the centre.  This resulted in
a galaxy number density ($n_{{\rm PSC}z}$) grid of total side length 32
$h^{-1}$ Mpc, far enough to enclose the local supercluster.  We then
compared all cubes 4 cells on a side from the simulations to the galaxy
density grid.  The comparison was made by minimizing the sum of
$(\rho-n_{{\rm PSC}z})^2$ over nine 16-$h^{-1}\!$ Mpc cubes: eight from
dividing the 32 $h^{-1}$ Mpc cube into octants, and one more in the centre.
We then shifted the best-fitting region to the centre of the 256 $h^{-1}\!$ Mpc
set of initial conditions before calculating the lower-box size initial
conditions.  Table \ref{init} shows these densities for the best fit.
Unfortunately, this procedure did not result in structures with obvious
visible similarity to the Local Group, but the statistical similarity is
reassuring.

Although similar, the inner regions of the four simulations were not
identical after evolving them to the present epoch.  Large-scale power
caused bulk motion in the central region, moving it away and distorting it
slightly from its original position.  To assess this effect, we
approximated it to first order with the sum of a translation and a linear
transformation, an asymmetric tensor which in general could include shear
and rotation.  We did not use this transformation in any analysis except in
evaluating the similarity of the four simulations.  The transformation can
be written as
\begin{equation}
  r_i^{(2)} = C_{ij}r^{(1)}_j + s_i,
  \label{transdef}
\end{equation}
where $r_i^{(m)}$ is the $i$th coordinate in the central region of
simulation $m$, $C_{ij}$ is a deformation tensor, and $s_i$ is a
translation vector.

\cenpic

Figure \ref{cenpic} shows particles initially in the central 16 $h^{-1}$
Mpc box from all simulations in the present epoch.  The region from the 32
$h^{-1}\!$ Mpc simulation is untransformed, with the other three regions
translated and deformed to fit best on to it.  To calculate the shift and
deformation, we compared the present-epoch positions of particles which had
occupied the same place in the initial conditions; i.e., with the same
Lagrangian positions.  Since mass resolution changes across simulations, it
was necessary to average together positions of particles (in sets of $2^3$,
$4^3$, or $8^3$) to compare positions in a smaller box size simulation to
those in a larger one.

We calculated the translation vector $s_i$ with $\langle r^{(2)}_i -
r^{(1)}_i\rangle$, where the angular brackets denote an average over
particles.  As for the deformation tensor $C_{ij}$, consider the quantity
$\langle (r_i^{(2)}-s_i)r^{(1)}_k\rangle$.  Assuming that Eqn.\
(\ref{transdef}) holds, this equals $C_{ij}\langle r^{(1)}_j
r^{(1)}_k\rangle$.  Thus, where $B_{ik} = \langle
(r_i^{(2)}-s_i)r^{(1)}_k\rangle$ and $A_{jk} = \langle r^{(1)}_j
r^{(1)}_k\rangle$,
\begin{equation}
  C_{ij} = B_{ik}(A^{-1})_{kj}.
  \label{transsol}
\end{equation}

The translation vectors $s_{i}$ from the 64, 128, and 256 $h^{-1}\!$ Mpc
simulations to the 32 had magnitudes 2.78, 4.43, and 5.92 $h^{-1}\!$ Mpc,
respectively.  The deformation tensors $C_{ij}$ were close to identity
matrices, as one would hope; the diagonal elements were all between 0.89
and 1.06 except for one outlier at 0.81, and the off-diagonal elements had
magnitudes below (mostly well below) 0.06.  With the first-order
adjustment, the agreement is still not perfect, but we expect the shear
from large-scale waves outside the inner box to change slightly over its
length; moreover, slight differences are likely to amplify when non-linearly
evolved.

\subsection{Halo finding}

To compare to the observed galaxy power spectrum, it is necessary first to
find haloes in the set of dark-matter particles returned by the
simulations.  Although they do reasonable jobs, no halo-finding
algorithm (HFA) is perfect.  In some analyses of N-body simulations,
surprisingly little discussion is given of the choice of HFA.

The first step of most HFAs is to estimate the density, a quantity which is
not obviously defined given only a set of particles.  DENMAX (Bertschinger
\& Gelb 1991) uses an Eulerian approach, calculating the density on a fine
mesh by smoothing each particle with a Gaussian of a fixed size, called the
smoothing length, $r_{\rm smoo}$, on an effectively infinitely fine mesh.
A Lagrangian approach (HOP, Eisenstein \& Hut 1998) uses a fixed number
$N_{\rm dens}$ of nearest-neighbor particles to estimate the density at the
position of each particle, and also uses a few other parameters.  DENMAX
has a fixed spatial resolution, while HOP effectively has a fixed mass
resolution.  The results of both methods are strongly dependent on their
free parameters, $r_{\rm smoo}$ or $N_{\rm dens}$.

Although DENMAX takes much more time to run than HOP, we ended up using a
variant of DENMAX.  We found that DENMAX is capable of finding smaller
haloes than HOP, down to about ten particles.  DENMAX works by moving
particles along density gradients until they are at a maximum.  It then
uses a `Friends-of-Friends' algorithm, finding clusters of moved
particles closer than a small linking length ($1/1024$ times the box size)
to each other.  The last step is to `unbind' iteratively any particles 
whose energies exceed the escape energy from their haloes.  The output of
DENMAX is a list of haloes with their masses (number of bound particles),
and their position and velocity centroids.

In DENMAX, a large smoothing length smears out close pairs of haloes, while
a small smoothing length, with its higher density threshold, fails to
include the outskirts of haloes in their mass, and also misses isolated,
less-dense haloes.  Gelb \& Bertschinger (1994) discuss some effects of
DENMAX resolution.  Empirical tests have indicated that setting $r_{\rm
smoo}=1/5$, in units of the mean interparticle separation, yields a halo
mass spectrum similar to that given by the Press--Schechter (1974)
formalism, a useful, though not omniscient, guide.  This choice of $r_{\rm
smoo}$ makes some sense theoretically, too, since the spherical collapse
model (Gunn \& Gott 1972) predicts that a virialized object is $\delta
\approx 180$ times denser than the background in a standard flat ($\Omega_m
= 1$) cosmology, and somewhat higher than that in a $\Lambda$CDM cosmology.
Fiducially, regions of overdensity $200$ and above are virialized,
corresponding to a smoothing length of $1/200^\frac13$, about $1/5$.
\dmsqr

We applied DENMAX with the canonical smoothing length to the results of
each simulation, and calculated halo power spectra.  The halo-finding
resolution we obtained was rather poor; there was a small-scale downturn in
each power spectrum starting at significantly larger scales than the
simulation's softening length.  We therefore tried halving the DENMAX
smoothing length to $r_{\rm smoo}=1/10$, which succeeded in extending the
power law in the correlation function to smaller scales by about a factor
of two. The smaller smoothing length evidently picked out subhaloes which
the canonical smoothing length had merged together.  A small smoothing
length is desirable in detecting subhaloes within a halo, since in a halo,
the spatial scales involved are smaller, and the background density is
higher.

We wanted to use higher resolution in higher-density regions without
forsaking the advantages of the canonical $r_{\rm smoo}$ in lower-density
regions.  We therefore used an algorithm which we call DENMAX$^2$, in which
DENMAX is run as normal with the canonical $r_{\rm smoo} = 1/5$, but then
is applied to each returned halo separately with $r_{\rm smoo} = 1/10$.
Although this factor of two is rather arbitrary, it put the DENMAX$^2$
$r_{\rm smoo}$ close to, but still above, the softening length in the 32
$h^{-1}$ Mpc simulation, and it was a convenient choice with so many other
factors of two being used, e.g.\ between the box sizes.  We could have kept
the DENMAX$^2$ $r_{\rm smoo}$ the same in physical units between the
boxsizes, but instead fixed it in interparticle units.  This was for a
couple of reasons: we wanted to hold constant the ratio of the highest and
lowest reliably measured wavenumber; also, in the larger-boxsize
simulations, a fixed smoothing length in physical units would be tiny in
interparticle units, and would thus encounter undesirably high Poisson noise.

Figure \ref{dmsqr} shows the effect of extra DENMAX$^2$ substructure on the
best-fitting (defined below) halo correlation functions in the 32
$h^{-1}\!$ Mpc simulation; it extends the power law to scales about half as
large, as one would expect from the halving of $r_{\rm smoo}$.  The results
are promising, but the choices of $r_{\rm smoo}$ for DENMAX and DENMAX$^2$
are still arbitrary, indicating the desirability of a HFA without such free
parameters.

From the list of DENMAX and DENMAX$^2$ haloes, we had to pick a subset
which we thought could represent PSC$z$ galaxies.  Since DENMAX returns the
mass of each halo, this was an obvious property to use to characterize
haloes.  However, the mass of the largest DENMAX$^2$ subhalo is necessarily
less than that of its parent DENMAX halo, both because it contains a subset
of the parent halo's particles, and because DENMAX$^2$'s smaller $r_{\rm
rsmoo}$ detects smaller structures.  We thus could not compare DENMAX mass
and DENMAX$^2$ mass directly.  We could have tried to `correct' DENMAX$^2$
masses to DENMAX levels, which are probably more physically meaningful,
since they approximate the virial mass.  For example, we could have
compared the DENMAX masses of each halo split by DENMAX$^2$ to the
DENMAX$^2$ mass of its largest subhalo.  However, it was occasionally
unclear which subhalo represented the original halo.  Also, it is not
obvious that the relationship between DENMAX and DENMAX$^2$ masses would be
the same for satellite subhaloes and main subhaloes.

We felt it was both simpler and more consistent to fit using another
quantifier.  Figure \ref{mvrho} shows that the central density of a halo,
estimated by counting the number of particles within a fixed radius
$r_\rho$ of the halo's centre of mass as returned by DENMAX, is
well-correlated to its DENMAX mass.  Also, central density doesnot unfairly
handicap dense but (as detected by DENMAX) unmassive haloes in their quest
for inclusion in the halo list.  For the 32 $h^{-1}\!$ Mpc simulation, we
used $r_\rho = 20$ $h^{-1}\!$ kpc.  At twice the softening length, this was
about the smallest scale at which we could expect to obtain a meaningful
density estimate.  Unfortunately, we could not use the same $r_\rho$ for
the larger simulations, because the mass resolution became too poor,
causing the density estimate within a small region to be dominated by
Poisson noise.  So, we simply increased $r_\rho$ in proportion to the box
size of the simulation.

Another quantity we could have used to characterize haloes is maximum
circular velocity $v_c$.  Figure \ref{vc} shows a comparison of central
density to $v_c$ for DENMAX haloes; the relationship is rather tight.  To
calculate $v_c$, we used the procedure employed by the Bound Density Maxima
HFA (Klypin et al. 1997).  This procedure, described at
\url{http://astro.nmsu.edu/~aklypin/PM/pmcode/node6.html}, yields $v_c =
\sqrt{GM(r)/r}$, evaluated at the radius of the first concentric shell
about the halo where the overdensity inside dips below 200, and also checks
its spherically averaged density profile to thwart structures hoping to
inflate $v_c$ for nearby haloes.  The maximum circular velocity is a useful
quantity because it can seemingly be related to observed galactic circular
velocities.  However, it is fallible, relying as it does on spher averaged
density profiles.  In any case, the relationship between the two quantities
is quite tight, so they are roughly interchangeable.

\mvrho
\vc

We thus picked subsets of the halo list according to a lower density cutoff
$\rho_{c,min}$; it is physically reasonable that the central density in a
halo must exceed some threshold to house an observed galaxy.  We first
imposed the density cutoff on the list of DENMAX haloes.  If application of
DENMAX$^2$ split a halo in the resulting subset into subhaloes, we imposed
the density cutoff on each of the subhaloes.  If none of these subhaloes
exceeded the density cutoff, the original halo stayed in the list;
otherwise, any dense-enough subhaloes (which almost always include the
original) replaced the original halo in the list.

We then calculated the haloes' power spectrum by binning halo pairs by
their separation, and then submitting the resulting correlation function to
an FFT (actually FFTLog, Hamilton 2000).  We also tried two other methods.
In the first, we calculated the density on a mesh, using the Nearest Grid
Point interpolation scheme, and found the power spectrum using a 3D FFT.
To get to the smallest scales, Klypin (private communication, 2001) pointed
out that if one divides a box of particles into octants (or some other
number of equal parts), and overlays all of the octants on each other,
periodic boundary conditions will still be satisfied, and the power
spectrum of the condensed box should be the same as that of the larger box.
This approach did work rather well, but there were small discrepancies
between power spectra from different octant overlayings, and it was not
obvious how to combine them.  Another method we tried used an
unequally spaced FFT (Beylkin 1995), which uses multiresolution analysis
(wavelets) to calculate the exact FFT of a set of delta functions in mass
(i.e.\ particles), but a sufficiently large unequally spaced FFT required
more memory than was convenient.  These other methods agreed well with the
correlation function technique, but we ended up using the correlation
function technique because it is possible to calculate the exact
correlation function of a relatively small number of haloes quickly with no
resolution limit.  Also, this technique is not subject to the vagaries of
window functions which exist for standard FFTs.

While it is more direct to calculate a correlation function than a power
spectrum from a simulation, the opposite is true for a redshift survey such
as PSC$z$.  This is because the power spectrum in directions transverse to
the line of sight in a redshift survey is unaffected by redshift
distortions, which is not true for the correlation function.  So, in
comparing simulations to observations, it was always necessary to translate
one set of data into the same space (either real or Fourier) as the other.
Empirically, the correlation function varied more dramatically on the
small-scale end with the density cutoff $\rho_{c,min}$, and also it is
easier to interpret directly in terms of physical pairs of haloes, so we
used the correlation function for fitting.  It would have been better in
principle to use the errors in the power spectrum, since they are more
directly measured from PSC$z$.  However, HT have not at present found a
positive-definite covariance matrix for the PSC$z$ power spectrum.  Thus any
comparison we make to simulations would not be completely rigorous anyway.
We were still able to estimate the goodness of fit by ignoring
cross-correlations among data points (just using HT's error bars), making a
`pseudo-$\chi^2$' ($\tilde{\chi}^2$) statistic.  Where $\xi$ denotes the
correlation function, and $b$ is the box size of a simulation, we included
$\xi(r)$'s with $r$ between $b/256$ and $b/2$; with our bins $r_i$ varying
as they did by a factor of $\sqrt{2}$, this included 14 points in the
fit. We calculated $\tilde{\chi}^2 = \sum_{i} \left(\frac{\xi(r_i)-\xi_{\rm
PSCz}(r_i)} {\sigma_{{\rm PSC}z}(r_i)}\right)^2$, where $\xi_{{\rm PSC}z}(r)$
is the PSC$z$ correlation function at $r$, and $\sigma_{{\rm PSC}z}(r)$
is the error in $\xi_{{\rm PSC}z}(r) = \left|\frac{\xi_+ - \xi_-}2\right|$,
the average of the upper and lower error bars as reported by HT.
$\xi_{\rm PSCz}$, $\xi_-$, and $\xi_+$ were logarithmically (or linearly,
if adjacent data points straddled zero) interpolated if necessary.

\pwrfit
\pwrfittot
\bias
\pseudochi
\pseudochin
\pseudochit

\section{Results}\label{res}
Figure \ref{pwrfit} shows the best individually-fit power spectra from each
simulation.  The resulting bias factors (see next paragraph) appear in
Fig.\ \ref{bias}, and the $\tilde{\chi}^2$ curves for the fits appear in
Fig.\ \ref{pseudochi}.  Figure \ref{pseudochin} shows the $\tilde{\chi}^2$
curves as a function of halo number density.  Figures \ref{pwrfittot} and
\ref{pseudochit} show an alternative collective fit, still varying a
central density cutoff, but constraining the halo number density to be the
same in the smallest three simulations; see \S \ref{disc} for further
discussion.  With the first of these fits, we have reproduced the form of
the PSC$z$ power spectrum on scales $k\la 40$ $h$ Mpc$^{-1}$, or $r \ga
0.05 h^{-1}\!$ Mpc.  The softening length was $0.01 h^{-1}\!$ Mpc, so this
is not much worse than one would hope for given that haloes are
many-particle, extended objects which necessarily exclude each other on
scales comparable to their radius.  Correlation functions and power spectra
appear with their theoretical error bars in Figs.\ \ref{individ} and
\ref{individp}.  The correlation functions turn down at large scales
because waves start to damp out as one approaches half the box size.  These
figures also show the dark matter correlation functions from each
simulation; see Fig.\ \ref{bias} for a clearer view of their relationships.
The error bars in the correlation function were calculated by splitting the
simulation volume into octants and calculating correlation functions in
each.  The error at $\xi(k)$ is then $\sqrt{[{\rm Var}(\xi_i(k))]/8}$,
where $i$ runs over all octants.  The same technique was used to calculate
power spectrum error bars.

\individ
\individp

\begin{table*}
\caption{\small Fit Information.}
\label{fits}
\begin{tabular}{lllll}
  Box size $b$ [$h^{-1}\!$ Mpc] & 32 & 64 & 128 & 256\\
  Particle mass [$10^{8} h^{-1}\!$ M$_\odot$] & 1.8 & 15 & 120 & 940\\
  Halo mass $m_b$ [\# particles] & 764 $\pm$ 322 & 269 $\pm$ 121 & 71 $\pm$ 31 & 17 $\pm$ 8\\
  Halo mass ratio $m_b/m_{2b}$ & 2.8 $\pm$ 1.8 & 3.8 $\pm$ 2.4 & 4.2 $\pm$ 2.7\\
  Physical halo mass [$10^{11} h^{-1}\!$ M$_\odot$] & $1.4\pm0.6$, & $4.0\pm1.8$ & $8.4\pm3.7$ & $16\pm8$\\
  Circular velocity $v_c$ [km/s] & 72 $\pm$ 28 & 98 $\pm$ 24 & 112 $\pm$ 14 & 120 $\pm$ 13\\
  Halo number density [$h^{3}$ Mpc$^{-3}$] & 0.0244 & 0.0175 & 0.00873 & 0.0061\\
  $\rho_{c,min}$ [\# particles] & 122 & 66 & 22 & 6\\
  
\end{tabular}
\end{table*}

Figure \ref{bias} shows a plot of the bias factor $b(k) = \sqrt{P_{\rm
haloes}(k)/P_{\rm dm}(k)}$, a measure of the difference between the galaxy
and dark matter power spectra, for the four simulations.  The small-scale
downturns are caused by the resolution of the halo-finding algorithm in
each spectrum.  The shape of the bias function is similar to what Kravtsov
\& Klypin (1999) found; the dependence of bias on scale must be similar to
this to achieve a power-law form in the galaxy power spectrum, given the
inflection in the dark matter power spectrum caused by the onset of
non-linearity, and the turnover back at smaller scales due to
virialization.

\begin{table*}
  \caption{ \small Splitting of the same haloes with decreasing box size
  (and thus decreasing DENMAX$^2$ smoothing length).  The halo number is
  the mass rank in the 32 $h^{-1}$ Mpc simulation DENMAX list, and $b$
  denotes the box size, in $h^{-1}$ Mpc.}
  \label{split}
  \begin{tabular}{lllll}
    $b$ & Halo 1 & Halo 3 & Halo 25 & Halo 26\\
    32 & 55 & 55 & 29 & 25\\
    64 & 19 & 20 & 9 & 10\\
    128 & 4 & 3 & 6 & 3\\
    256 & 2 & 1 & 1 & 1\\
\end{tabular}
\end{table*}

Table \ref{fits} shows the central density cutoffs of the best fits, along
with the number densities of the best-fitting populations, and effective
cutoffs in halo mass (from Fig.\ \ref{mvrho}) and maximum circular velocity
(from Fig.\ \ref{vc}).  The quoted errors are the standard deviations in
mass or $v_c$ of haloes with $\rho_c$ within 1/10 of a dex of
$\rho_{c,min}$.  We ignore errors arising from the goodness-of-fit, which
we did not include since our $\tilde{\chi}^2$ estimate is not rigorous.
These errors could be sizable, though, particularly in the 256 $h^{-1}\!$
Mpc simulation.  The minimum at 6 particles in Fig.\ \ref{pseudochi} for
the 256 $h^{-1}\!$ Mpc simulation's haloes is quite shallow.  The
$\tilde{\chi}^2$ value including all detected haloes ($\rho_{c,min} = 0$)
was only slightly greater than at $\rho_{c,min} = 6$.  Using either cutoff,
most of the haloes are right at the detection limit (a DENMAX mass of 10
particles), so it is quite possible that the mass resolution in this
simulation is insufficient to pick up the truly best-fitting population of
haloes.

\subsection{Discussion}
Figure \ref{pwrfit} shows good fits to the PSC$z$ power spectrum for each
simulation, but the question remains: could these haloes from four
different simulations represent the same set of haloes?  Encouragingly,
when translated into maximum circular velocity, the cutoff intervals mostly
overlap with each other.  The only disjoint error intervals are between the
256 and the 32 $h^{-1}$ Mpc simulations, and once again, the errors in the
256 $h^{-1}$ Mpc case are probably underestimated.  However, we do not expect
them to be exactly the same populations, since in a smaller box, the higher
mass and DENMAX$^2$ spatial resolutions produce more haloes with small
separations, many of which could join together in a lower-resolution
simulation; this fact is evident in the increased small-scale range of the
correlation functions in each simulation.  Still, there are two apparent
discrepancies which should be understood in Table \ref{fits}: in halo
number density and in implied mass cutoff.

One possible explanation for the different-looking halo populations is
cosmic variance; i.e.\ that our chosen set of initial conditions was funny
in some way.  For example, because of periodic boundary conditions, the 32
$h^{-1}$ Mpc simulation is not big enough to contain the local supercluster
from the PSC$z$ cube we used to pick the initial conditions.  We measured the
correlation functions of haloes in the central 16 $h^{-1}$ Mpc cube of each
simulation, first applying the shifts and deformation tensors necessary to
put the larger simulations on top of the 32 $h^{-1}$ Mpc simulation.
Figure \ref{cencrl} shows the results of this test, for both the
individually-fit and number density-constrained central density cutoffs.
The central box is evidently undercorrelated relative to the larger boxes,
but the correlation functions from different simulations seem consistent
with each other, even when the sets of haloes are constrained to have the
same number density.

\cencrl
\halocmp
\halofig

To try to understand the behavior of DENMAX mass and halo number density
across the simulations, we identified a few haloes by eye in the central
regions of the four simulations.  We looked for large haloes near each
other across the simulations, and then visualized them with the {\tt
points} program written by Michael Blanton, at
\url{http://physics.nyu.edu/~mb144/graphics.html}.  Identifying
the same halo in each simulation was complicated by the bulk motion and
distortion of the central regions, as discussed at the end of \S \ref{sim}.
We found four obvious haloes; the numbers of particles which comprised them
in the 256 $h^{-1}\!$ Mpc simulation were 636, 501, 100, and 89.  Figure
\ref{halofig} shows one of them (`Halo 3' from Table \ref{split}), and how
it was split by DENMAX$^2$.  While the results across simulations are
similar, some subhaloes appear differently in different simulations, and
some of them are questionably existent.  If subhaloes have orbited their
parent halo a few times, they might not be in the same place across all
simulations, since the simulations do quite well in tracking orbits, but
not necessarily phases along them.  Figure \ref{halocmp} shows density
profiles for the four haloes identified across the simulations, which agree
quite well.  In the next two sections, we will discuss what we learned from
these haloes.

\subsubsection{Number Density}\label{disc}

From Table \ref{fits}, the number density of best-fitting haloes changes by a
factor approaching two between adjacent simulations.  To gauge the severity
of this potential problem, we investigated the effect of forcing the number
density of haloes in each box to match.  Figure \ref{pseudochin} shows
$\tilde{\chi}^2$ as a function of halo number density.  Excluding the 256
$h^{-1}\!$ Mpc simulation because it was not clear that the desired set of
haloes was well-resolved, and adding together $\tilde{\chi}^2$ values from
other simulations with equal weight, we obtained a total $\tilde{\chi}^2$
curve, which appears in Fig.\ \ref{pseudochit}.  The best-fitting number
density from Fig.\ \ref{pseudochit} was 0.0182 $h^{-3}$ Mpc$^3$, which
matches the number density of PSC$z$ galaxies that are about 20 $h^{-1}$ Mpc
from the Milky Way.  Figure \ref{pwrfittot} shows the resulting power
spectra, which are not particularly consistent, either with each other or
with PSC$z$.  The number density constraint allows smaller haloes into the
128 $h^{-1}\!$ Mpc set, which lowers the power spectrum; it has the
opposite effect on the 32 $h^{-1}\!$ Mpc set, removing smaller haloes and
lifting the power spectrum.

However, it is not obvious that we should expect the same number density of
haloes when resolution changes.  Suppose we have a set of $N$ haloes in a
box of volume $V$.  To measure the correlation function, we bin pairs of
haloes by distance $r_i$ to obtain $P(r_i)$, the number of pairs in bin
$i$, with volume $V(r_i)$.  The value of the correlation function
$\xi(r_i)$ in bin $i$ is then given by:
\begin{equation}
  1 + \xi(r_i) = \frac{V}{V(r_i)}\frac{P(r_i)}{N^2}.
  \label{xidef}
\end{equation}
The RHS expresses the number of halo pairs in a bin, normalized by dividing
by three quantities: the volume of the bin, the number of haloes, and the
mean halo number density.  Now suppose that each of these haloes in fact
consists of two subhaloes which become resolved when the resolution
increases, and that all separations between these subhaloes are much
smaller than the $r_i$'s.  Each pair turns into four, and $N$ doubles.  The
new correlation function $\xi^\prime$ over the previous range of $r_i$'s is
given by
\begin{equation}
1 + \xi^\prime(r_i) = \frac{V}{V(r_i)}\frac{4P(r_i)}{(2N)^2} = 1 + \xi(r_i),
  \label{iprime}
\end{equation}
and pairs also appear in new, smaller-scale bins, imparting to the
correlation function there a value possibly as large as the other
$\xi(r_i)$'s, depending on the bin size and distribution of subhalo
distances.  So in this simple case, when a higher spatial resolution
reveals more substructure, the halo number density increases without
changing the correlation function except in new, smaller-scale bins.

Might this model resemble our simulations?  Table \ref{split} shows how the
number of subhaloes uncovered by DENMAX$^2$ changes with box size for the
four haloes we identified.  This table includes all subhaloes, not merely
the ones which make the central density cutoff.  Also, these are relatively
large parent haloes, more likely to have substructure than smaller ones.
This seems to be a reasonable explanation for the change in halo number
density across the simulations.

\subsubsection{Mass cutoff}

Where $m_b$ is the DENMAX mass (in number of particles) of a halo in a
simulation of box size $b$, $m_b/m_{2b}$ should ideally be 8, the same
factor by which the mass resolution changes.  The values of $m_b/m_{2b}$
for the best fits in Table \ref{fits} fall well short of this.  For our
four-halo sample, $m_{32}/m_{64} = 6.3 \pm 0.8$, $m_{64}/m_{128} = 6.1 \pm
0.6$, and $m_{256}/m_{128} = 7.3 \pm 0.5$, somewhere between the ratios in
Table \ref{fits} and the expected value of 8.  The shrinkage of a halo with
increased mass resolution is obvious in Fig. \ref{halofig}, and comes
primarily from the reduction in $r_{\rm smoo}$ in physical units as
particle mass decreases.

\section{Conclusions}

We have shown that we can reproduce the clustering properties of
infrared-selected PSC$z$ galaxies fairly well in simulations for scales $k
\la 40$ $h$ Mpc$^{-1}$, near our resolution limit, by imposing a
central density cutoff on haloes.  This central density cutoff implies a
rough dark matter mass cutoff in PSC$z$ galaxies of $10^{11-12}$
M$_\odot$, and a cutoff in maximum circular velocity of roughly 100 km/sec.
Thus, it appears that dark matter physics alone is sufficient to describe
the distribution of PSC$z$ galaxies on the scales we probed.  It is doubtful
that a there is a strict central density cutoff in haloes which house PSC$z$
galaxies in nature, but it seems that central density is a still a decent
indicator of the hospitality of haloes toward nascent galaxies.  While the
fits look good, their full statistical assessment awaits a covariance
matrix for the PSC$z$ power spectrum.

We have also found that the best-fitting halo populations from the four
simulations are probably consistent with each other.  Their circular
velocities are roughly consistent.  On the other hand, their number
densities and mass cutoffs do increase systematically with spatial
resolution, but we do not believe that these discrepancies indicate that
the populations are necessarily inconsistent.  One might in fact expect the
halo number density to increase when higher resolution reveals more
substructure, without affecting the correlation function.  Also, the
varying mass cutoff is at least partially an artifact of our halo-finding
algorithm.  We are currently considering new methods to identify haloes
with fewer (perhaps even zero) free parameters, which we hope will coax the
tips more confidently from their simulated dark-matter icebergs.

\section{Acknowledgments}

This work was supported by NASA ATP grant NAG5-10763, and a grant on the
Origin 2000 system by the National Computational Science Alliance.  We
thank an anonymous referee for helpful and insightful comments and
suggestions.  The work also advanced pleasantly at the Aspen Center for
Physics.

\end{document}